\documentclass{aa}  
\usepackage{natbib}
\usepackage[breaklinks,colorlinks,linkcolor=blue,citecolor=blue,urlcolor=blue]{hyperref}
\bibpunct{(}{)}{;}{a}{}{,} 
\usepackage{graphicx}
\usepackage{txfonts}
\begin{document}

\title{Existence of Tidal Tails for the Globular Cluster NGC 5824}


\author{
	Yong Yang \inst{1,2}
	\and
	Jing-Kun Zhao \inst{1} 
	\and
	Miho N. Ishigaki \inst{3,4,5}
	\and
	Masashi Chiba \inst{4}
	\and
	Cheng-Qun Yang \inst{6}
	\and
	Xiang-Xiang Xue \inst{1}
	\and
	Xian-Hao Ye \inst{1,2}
	\and
	Gang Zhao \inst{1,2}
}

\institute{
	CAS Key Laboratory of Optical Astronomy, National Astronomical Observatories, Chinese Academy of Sciences, Beijing 100101, People's Republic of China\\
	\email{zjk@bao.ac.cn}
	\and
	School of Astronomy and Space Science, University of Chinese Academy of Sciences, Beijing 100049, People's Republic of China
	\and
	National Astronomical Observatory of Japan, 2-21-1 Osawa, Mitaka, Tokyo 181-8588, Japan
	\and
	Astronomical Institute, Tohoku University, 6-3, Aramaki, Aoba-ku, Sendai, Miyagi 980-8578, Japan
	\and
	Kavli Institute for the Physics and Mathematics of the Universe (WPI), The University of Tokyo Institutes for Advanced Study, The University of Tokyo, 5-1-5 Kashiwanoha, Kashiwa, Chiba 277-8583, Japan
	\and
	Key Laboratory for Research in Galaxies and Cosmology, Shanghai Astronomical Observatory, 80 Nandan Road, Shanghai 200030, People's Republic of China
}



\abstract
{Several dynamically cold streams have been associated with certain globular clusters (GCs) based on orbital energies and angular momenta. Some of these streams are surprisingly far apart from their progenitors and one such pair is Triangulum and NGC 5824. Triangulum can be considered as a piece of NGC 5824 leading tail since the cluster's future orbit matches with the stream's track well. The existence of the leading tail for NGC 5824 is the motivation behind the search for its trailing tail.} 
{Our goal is to confirm the connection between Triangulum and NGC 5824 and seek the trailing tail of the cluster.}
{The selection of member stars of Triangulum is made through various cuts in metallicity, proper motions (PMs), radial velocity and color-magnitude diagram (CMD). The selected members are compared in phase space to a mock stream which models the disruption of NGC 5824. We then try to detect the trailing tail of the cluster based on a modified matched-filter technique. Stars are assigned weights using their color differences from the cluster's locus in CMD. These weights are further scaled based on stars' departures from expected PMs of the model stream.}
{A total of 26 member stars for Triangulum are obtained and 16 of them are newly identified. These members are consistent with the mock stream in the phase space and their metalicity and position on the CMD are in good agreements with NGC 5824. By applying the matched-filter, a tenuous trailing tail of the cluster is detected, spanning $\sim$ 50$\degr$ long on sky. The signature matches with the mock stream's trajectory well. }
{Our results support that Triangulum stream acts as a part of the leading tail for NGC 5824. On the trailing side, we have detected a 50$\degr$ tail extended from the cluster. The existence of both leading and trailing tails for the GC NGC 5824 is verified.}


\maketitle
%

\section{Introduction}

Increasing amount of data from various revolutionary surveys are revealing mysteries of stellar streams in the Milky Way and providing unprecedented details of the Galactic halo \citep[e.g.,][]{2008ApJ...680..295B,2009ApJ...692L.113Z,2010ApJ...714..229L,2015MNRAS.449.1391B,2016MNRAS.463.1759B,2017ApJ...844..152L,2018ApJ...868..105Z,2018MNRAS.481.3442M,2019ApJ...886..154Y,2019ApJ...880...65Y,2020ApJ...904...61Z,2021ApJ...922..105Y,2021AJ....162..171Y,2021SCPMA..6439562Z}. Tidal streams extending from extant globular
clusters (GCs) are usually thin and dynamically cold \citep[e.g.,][]{2003AJ....126.2385O,2006ApJ...639L..17G,2019MNRAS.488.1535P,2019ApJ...884..174G}. Some narrow streams without explicit cores are generally also attributed to GC origins \citep[e.g.,][]{2009ApJ...693.1118G,2010ApJ...712..260K,2012ApJ...760L...6B,2014MNRAS.442L..85K,2018ApJ...862..114S,2018MNRAS.481.3442M}. The progenitors of most of those streams are still unknown but several streams have been recently associated with extant GCs \citep{2021ApJ...914..123I}.

The connections between $\omega$ Centauri and Fimbulthul \citep{2019NatAs...3..667I}, NGC 3201 and Gj\"{o}ll \citep{2021MNRAS.504.2727P}, and NGC 4590 and Fj\"{o}rm \citep{2019MNRAS.488.1535P} have been reported, which suggest that the associations between a stream and a GC, where the GC does not connect directly to the stream, are present in the Milky Way. By exploring the orbits, \citet{2021ApJ...909L..26B} further attributed 5 more streams to extant GCs (Table~1 therein), and one pair is Triangulum and NGC 5824. Triangulum stream was first detected by \citet{2012ApJ...760L...6B} with a matched-filter technique \citep{2002AJ....124..349R}. Thereafter, \citet{2013ApJ...765L..39M} kinematically discovered a part of the stream and provided 11 possible member stars. The stream is in the direction of M31 and M33 galaxies, and far apart from NGC 5824. However, the cluster's future orbit passes through the stream well, implying a connection between them \citep[Fig.~4 in][]{2021ApJ...909L..26B}. \citet{2022ApJ...928...30L} further confirmed this connection by comparing a model stream of NGC 5824 in phase space to the Triangulum member stars from \citet{2013ApJ...765L..39M}. Therefore, Triangulum stream could be treated as a piece of NGC 5824 leading tail.

Based on the picture that tidal tails are developed symmetrically around GCs \citep{2010MNRAS.401..105K}, the existence of leading tail for NGC 5824 motivates us to search for its trailing tail. In this work, we provide a confirmation of the connection between Triangulum and NGC 5824, which is similar to that of \citet{2022ApJ...928...30L} but with member stars that span a wider sky extent ($\sim$ 16$\degr$). We further apply a modified match-filter method \citep{2019ApJ...884..174G} to look for the trailing tail of NGC 5824. The paper is organized as follows. In Sect.~\ref{sec:data}, we introduce the data. In Sect.~\ref{sec:connection}, we show the selection of Triangulum member stars and compare them to a model stream of NGC 5824. The detection of the cluster's trailing tail is given in Sect.~\ref{sec:match-filter}. We present a discussion in Sect.~\ref{sec:discussion} and draw our conclusion in Sect.~\ref{sec:summary}.


\section{Data}
\label{sec:data}

We base our search on high-quality astrometric and photometric data provided by the $Gaia$ EDR3 \citep{2021A&A...649A...1G,2021A&A...649A...4L,2021A&A...649A...3R}, along with the spectroscopic data from the Sloan Extension for Galactic Understanding and Exploration \citep[SEGUE;][]{2009AJ....137.4377Y} and the Large Sky Area Multi-Object Fiber Spectroscopic Telescope \citep[LAMOST;][]{2012RAA....12.1197C,2006ChJAA...6..265Z,2012RAA....12..723Z,2015RAA....15.1089L} surveys.

\begin{figure*}
   	\includegraphics[width=\linewidth]{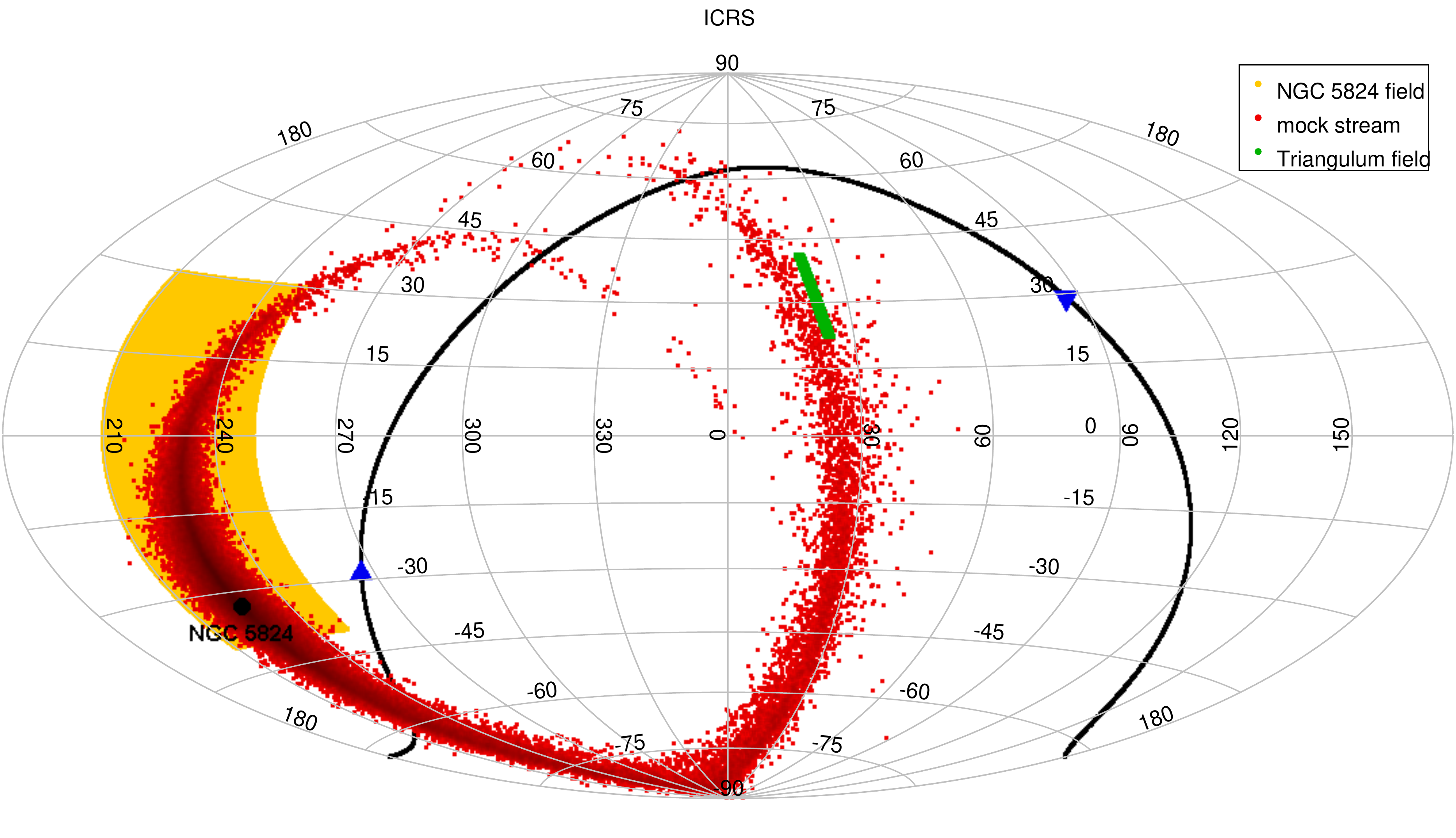}
   	\caption{Sky projections of the data (green and orange areas) and a mock stream (red dots). The black line represents the Galactic plane, and the blue (inverted) triangle represents the direction of Galactic (anti-) center. The black circle denotes the GC NGC 5824.}
   	\label{fig:icrs}
\end{figure*}

To obtain the individual members of Triangulum, we retrieve stars from the $Gaia$ EDR3 \texttt{gaia\_source} catalog overlapping with the stream region on the celestial sphere. The stream region is determined by limiting 22$\degr$ $<$ $\delta$ $<$ 41$\degr$ and moving $\delta = -4.4\alpha + 128.5$ by $\pm 1\degr$ along the $\alpha$ direction (green area in Fig.~\ref{fig:icrs}), where the equation was defined in \citet{2012ApJ...760L...6B} to describe the stream coordinates. Note that \citet{2012ApJ...760L...6B} traced Triangulum to $\delta \simeq 23\degr - 35\degr$, and \citet{2014ApJ...787...19M} extended the stream to further north $\delta \simeq 40\degr$. Our choice of $\delta$ extent is based on both of them. The zero-point correction in the parallax is implemented using the code provided by \citet{2021A&A...649A...4L}, which requires \texttt{astrometric\_params\_solved} $>$ 3. The corrections of G-band magnitude and BP/RP excess factor are applied as instructed in \citet{2021A&A...649A...3R}. In order to ensure good astrometric and photometric solutions, only stars with \texttt{ruwe} $<$ 1.4 and absolute corrected BP/RP excess factor smaller than 3 times the associated uncertainty \citep[see Sect.~9.4 in][]{2021A&A...649A...3R} are retained. Given that both of estimated distances of the stream in \citet{2012ApJ...760L...6B} and \citet{2013ApJ...765L..39M} are farther than 20 kpc, we remove foreground stars that satisfy the criterion $\varpi - 3\sigma_{\varpi} > 0.05$ mas. The remaining stars are cross-matched with SDSS/SEGUE DR16 \citep{2012ApJS..203...21A} and LAMOST DR8, by which the metallicity and heliocentric radial velocity are obtained. For stars that are common in both datasets, we adopt measurements from SEGUE because signal-to-noise ratios of spectra in SEGUE are mostly higher than those in LAMOST. 

The data for detecting trailing tail of NGC 5824 are also obtained from $Gaia$ EDR3. Stars within the sky box of 210$\degr$ $<$ $\alpha$ $<$ 250$\degr$ and -40$\degr$ $<$ $\delta$ $<$ 30$\degr$ are retrieved (orange area in Fig.~\ref{fig:icrs}) and reduced with the same procedures as above (including the foreground stars removing\footnote{Removing foreground stars within 20 kpc will not affect results since if the cluster's trailing tail exists, it would be farther than 30 kpc from the sun (see Fig.~\ref{fig:locus}).}). Since the spectroscopic surveys are unavailable in this sky region, only $Gaia$ data are used.

In Fig.~\ref{fig:icrs}, we show projections of the data (green and orange areas), along with a mock stream (red dots) which will be described in Sect.~\ref{subsec:mockstream}. The black line represents the Galactic plane, and the blue (inverted) triangle represents the direction of Galactic (anti-) center. It should be noted that the NGC 5824 field is exactly designed based on the mock stream. 

\section{Connection between Triangulum and NGC 5824}
\label{sec:connection}

\subsection{Triangulum Member Stars}
\label{subsec:members}

\begin{figure}
  	\includegraphics[width=0.9\columnwidth]{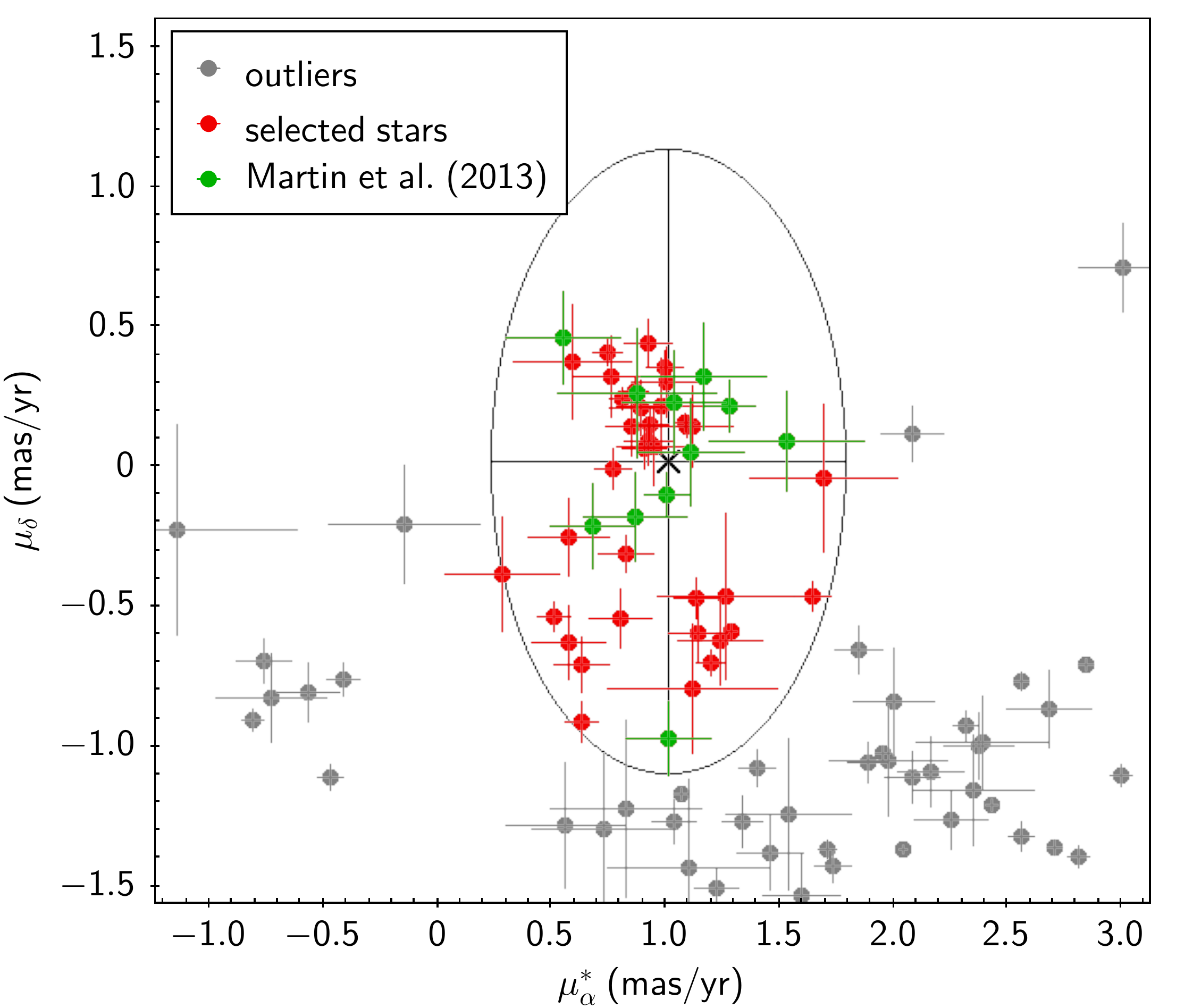}
  	\includegraphics[width=0.9\columnwidth]{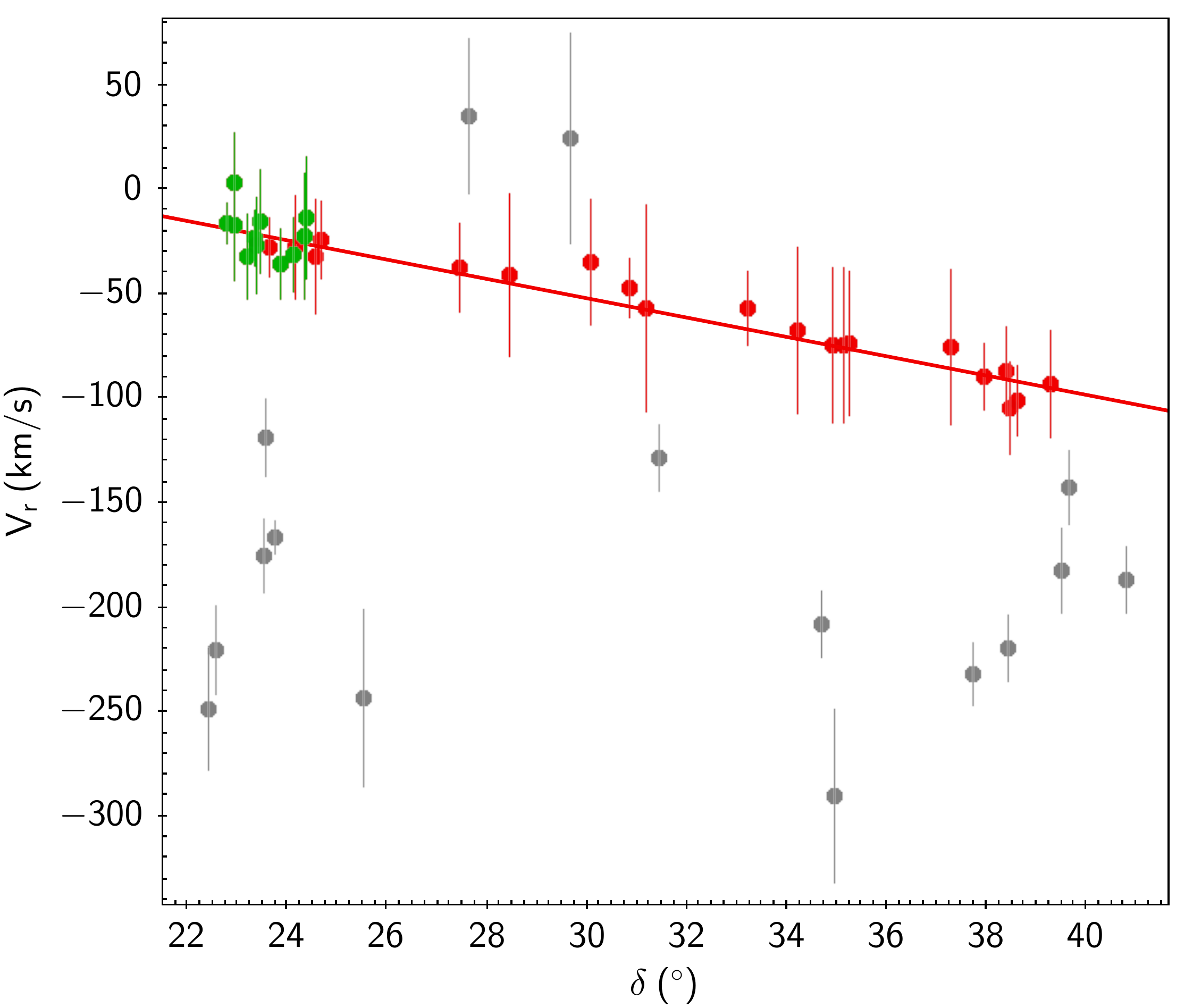}
  	\includegraphics[width=0.9\columnwidth]{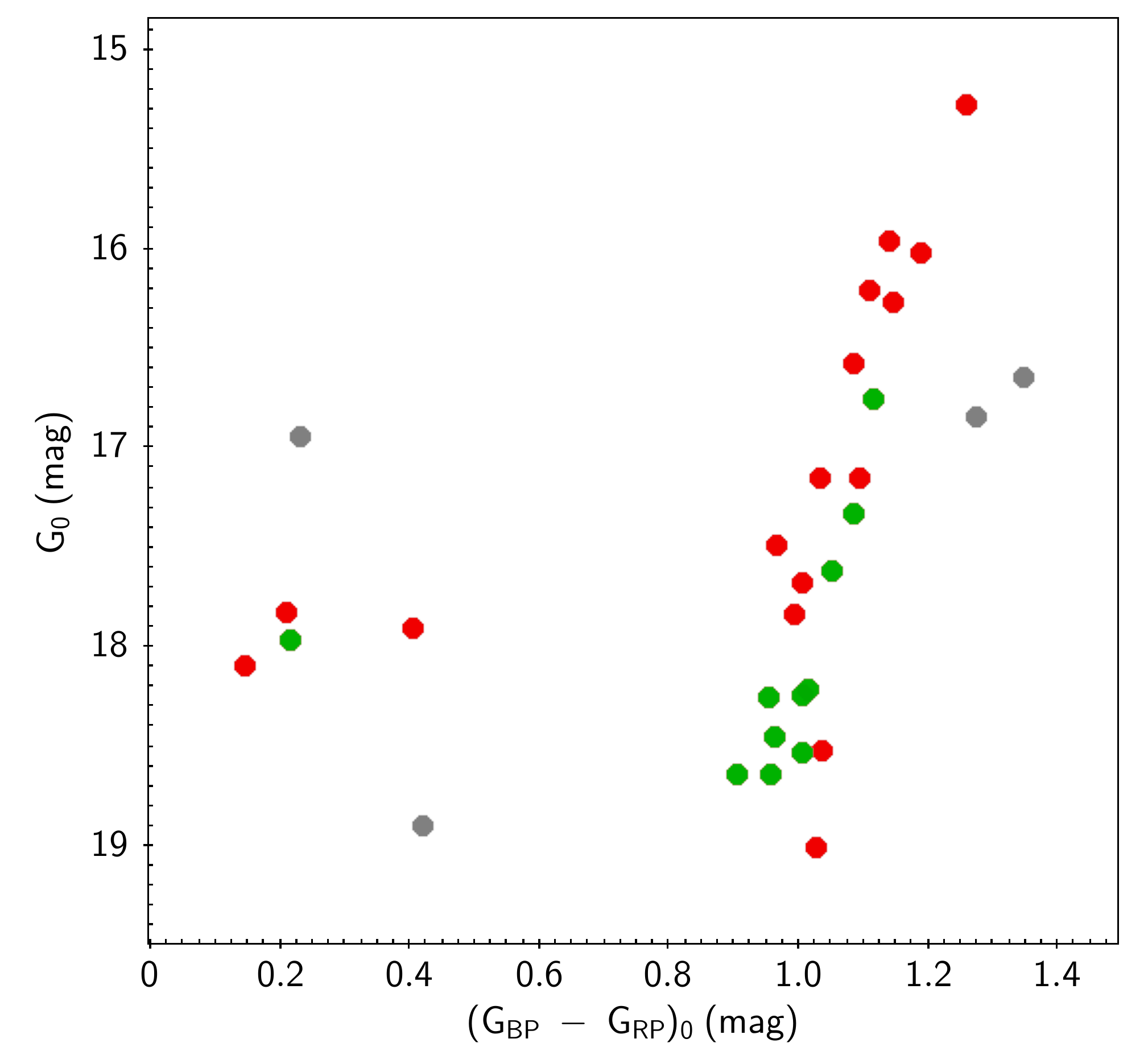}
  	\caption{The selections of Triangulum member stars. The gray dots represent rejected stars and the red ones represent the selected stars during each step. The member candidates identified by \citet{2013ApJ...765L..39M} are marked in the green points. The top panel shows the local region of the overdensity in PM space, where the ellipse is defined to select member candidates in this step. The middle panel shows stars in $\alpha$ - $V_r$  plane, where the error bars represent three times uncertainties of $V_r$ and the red line is a linear fit to the stream sequence. The bottom panel shows those candidates in CMD. }
  	\label{fig:selection}
\end{figure}

Cross-matching between $Gaia$ sources and spectroscopic data yields 1,968 stars. \citet{2012ApJ...760L...6B} presented an estimate of Triangulum's [Fe/H] to be $\sim$ -1.0 dex, while \citet{2013ApJ...765L..39M} contended a poorer metallicity $\simeq$ -2.2 dex for the stream. In order to obtain as many member stars as possible, we adopt [Fe/H] $<$ -1.0 dex as the selection criterion and are left with 451 candidates. After this cut, an overdensity can be seen clearly in proper motion (PM) space (top panel of Fig.~\ref{fig:selection}, only the local region around the overdensity is shown). We overplot the member candidates provided by \citet{2013ApJ...765L..39M} (cross-matched with $Gaia$ EDR3) and verify that this overdensity exactly corresponds to Triangulum stream. To pick out stream stars, we define a dispersion ellipse whose center and semi-axes are determined based on the known candidates from \citet{2013ApJ...765L..39M}. The center (1.014, 0.012) mas yr$^{-1}$ is the mean PM of the members in $\alpha$ and $\delta$, and the semi-axes (0.777, 1.116) mas yr$^{-1}$ are three times the PM dispersions in respective directions. 47 stars enclosed within the ellipse are selected. 

These stars are then plotted in $\delta$ - $V_r$ plane (middle panel of Fig.~\ref{fig:selection}) and a dominant monotonic sequence is clearly discernable. Generally, the radial velocities of a halo stream are supposed to change monotonically along coordinates as long as there is no turning point contained (like apogalacticon), such as Pal 5 \citep{2016ApJ...823..157I}, GD-1 \citep{2016ApJ...833...31B}, NGC 5466 \citep{2022MNRAS.513..853Y}, Hr\'{i}d and Gj\"{o}ll stream \citep{2021ApJ...914..123I}. Hence we consider that this dominant sequence should correspond to Triangulum stream. We fit a straight line to the sequence where weights are given by the uncertainties of $V_r$. The relation can be described with the equation $V_r = -4.6\delta + 86.5$. 31 stars with $V_r$ consistent to the fit in 3$\sigma$ range are retained. 

Finally, we reject 4 more outliers on the basis of color-magnitude diagram (CMD) and 27 member stars follow a typical GC isochrone (bottom panel of Fig.~\ref{fig:selection}). All sources here have been extinction-corrected using the \citet{1998ApJ...500..525S} maps as re-calibrated by \citet{2011ApJ...737..103S} with RV = 3.1, assuming $A_{G}/A_{V} = 0.83627$, $A_{BP}/A_{V} = 1.08337$, $A_{RP}/A_{V} = 0.63439$\footnote{These extinction ratios are listed on the Padova model site \url{http://stev.oapd.inaf.it/cgi-bin/cmd}.}. The detailed information of 27 member stars is summarized in Table~\ref{tab:members}. 

\begin{sidewaystable*}
  	\caption{Triangulum stream member stars.}
  	\label{tab:members}
  	\centering
  	\begin{tabular}{lcccccccccccccc}
 	 \hline\hline
 	 No. & $\alpha_{J2000}$ & $\delta_{J2000}$ &  $\mu_{\alpha}^*$ & $\sigma_{\mu_{\alpha}^*}$ & $\mu_{\delta}$ & $\sigma_{\mu_{\delta}}$ & $V_r$ & $\sigma_{V_r}$ & [Fe/H] & $\sigma_{\rm [Fe/H]}$ & $G$ & $G_{bp}$ & $G_{rp}$ & Survey \\
	   & (\degr) & (\degr) &  (mas yr$^{-1}$) &(mas yr$^{-1}$)& (mas yr$^{-1}$) &(mas yr$^{-1}$)& (km\,s$^{-1}$) &(km\,s$^{-1}$) & (dex) &(dex)& (mag) & (mag) & (mag) &   \\
  	\hline
  	1$^\star$ & 23.8285 & 22.8031 & 1.0098 & 0.0986 & -0.1037 & 0.0784 & -16.84 & 3.11 & -1.913 & 0.068 & 16.759 & 17.222 & 16.106 & SEGUE\\
  	2$^\star \times$ & 24.1829 & 22.9364 & 1.0127 & 0.1831 & -0.9748 & 0.1339 & 2.78 & 7.84 & -2.545 & 0.048 & 17.617 & 18.049 & 16.997 & SEGUE\\
  	3$^\star$ & 24.2157 & 22.9598 & 0.8790 & 0.3456 & 0.2562 & 0.2288 & -17.60 & 8.73 & -1.867 & 0.104 & 18.530 & 19.005 & 17.998 & SEGUE\\
  	4$^\star$ (BHB) & 23.2433 & 23.1934 & 0.6858 & 0.1844 & -0.2179 & 0.1546 & -32.59 & 6.76 & -2.121 & 0.078 & 17.967 & 18.031 & 17.815 & SEGUE\\
  	5$^\star$ & 24.1515 & 23.3639 & 1.2849 & 0.1128 & 0.2147 & 0.0938 & -23.76 & 4.55 & -2.348 & 0.114 & 17.335 & 17.803 & 16.718 & SEGUE\\
  	6$^\star$ & 23.9200 & 23.3903 & 0.5511 & 0.2549 & 0.4586 & 0.1641 & -26.84 & 7.61 & -2.128 & 0.114 & 18.251 & 18.672 & 17.667 & SEGUE\\
  	7$^\star$ & 23.8055 & 23.4783 & 1.1128 & 0.2314 & 0.0493 & 0.1889 & -16.15 & 8.19 & -2.371 & 0.056 & 18.457 & 18.891 & 17.927 & SEGUE\\
  	8 (BHB) & 24.5968 & 23.6575 & 1.1237 & 0.1755 & 0.1393 & 0.1465 & -28.03 & 4.77 & -2.278 & 0.152 & 17.834 & 17.903 & 17.692 & SEGUE\\
  	9$^\star$ & 23.7817 & 23.8800 & 1.0403 & 0.2248 & 0.2268 & 0.1843 & -36.43 & 5.51 & -2.383 & 0.095 & 18.256 & 18.669 & 17.716 & SEGUE\\
  	10$^\star$ & 23.9193 & 24.1185 & 0.8721 & 0.2242 & -0.1813 & 0.1614 & -31.52 & 5.94 & -1.953 & 0.128 & 18.220 & 18.705 & 17.689 & SEGUE\\
  	11 & 24.0593 & 24.1634 & 1.0918 & 0.0460 & 0.1548 & 0.0351 & -27.93 & 8.18 & -2.249 & 0.061 & 15.279 & 15.832 & 14.573 & LAMOST\\
  	12$^\star$ & 23.4145 & 24.3451 & 1.5329 & 0.3380 & 0.0838 & 0.1762 & -22.60 & 9.99 & -2.072 & 0.072 & 18.648 & 18.981 & 18.075 & SEGUE\\
  	13$^\star$ & 23.5792 & 24.3911 & 1.1712 & 0.2725 & 0.3187 & 0.1921 & -13.69 & 9.70 & -2.653 & 0.147 & 18.648 & 19.030 & 18.073 & SEGUE\\
  	14 & 23.5445 & 24.5692 & 1.0955 & 0.0584 & 0.1400 & 0.0363 & -32.73 & 9.18 & -1.962 & 0.135 & 16.025 & 16.537 & 15.347 & LAMOST\\
  	15 & 23.8850 & 24.7038 & 0.9478 & 0.1592 & 0.0680 & 0.1383 & -24.41 & 6.23 & -1.830 & 0.140 & 17.839 & 18.239 & 17.246 & SEGUE\\
  	16 & 22.3385 & 28.4684 & 0.9975 & 0.0801 & 0.3481 & 0.0569 & -41.53 & 12.97 & -2.311 & 0.111 & 16.581 & 17.045 & 15.961 & LAMOST\\
  	17 & 22.4622 & 30.0863 & 0.9254 & 0.1081 & 0.0873 & 0.0825 & -34.82 & 9.93 & -1.716 & 0.272 & 17.156 & 17.596 & 16.563 & LAMOST\\
  	18 & 22.3181 & 31.2075 & 0.9126 & 0.0968 & 0.0614 & 0.0716 & -57.01 & 16.48 & -2.143 & 0.241 & 17.152 & 17.616 & 16.521 & LAMOST\\
  	19 & 21.4457 & 34.2315 & 0.8154 & 0.0552 & 0.2357 & 0.0419 & -68.02 & 13.15 & -2.439 & 0.161 & 15.964 & 16.453 & 15.313 & LAMOST\\
  	20 & 20.8085 & 34.9178 & 0.7522 & 0.0638 & 0.4043 & 0.0443 & -75.26 & 12.41 & -2.056 & 0.136 & 16.213 & 16.680 & 15.569 & LAMOST\\
  	21 & 21.2461 & 35.1423 & 0.9246 & 0.1032 & 0.4356 & 0.0884 & -74.81 & 12.45 & -2.322 & 0.204 & 17.493 & 17.885 & 16.919 & LAMOST\\
  	22 & 21.3470 & 35.2527 & 0.8679 & 0.0563 & 0.2653 & 0.0435 & -74.42 & 11.61 & -2.141 & 0.211 & 16.273 & 16.759 & 15.613 & LAMOST\\
  	23 & 20.6703 & 37.3033 & 0.5967 & 0.2600 & 0.3732 & 0.2044 & -76.24 & 12.45 & -2.145 & 0.038 & 19.007 & 19.514 & 18.485 & SEGUE\\
  	24 (BHB) & 20.4665 & 37.9600 & 0.7622 & 0.1674 & 0.3160 & 0.1455 & -90.22 & 5.21 & -1.740 & 0.055 & 18.101 & 18.157 & 18.010 & SEGUE\\
  	25 & 20.6305 & 38.4133 & 0.9838 & 0.1966 & 0.2130 & 0.1741 & -87.51 & 7.20 & -2.042 & 0.089 & 18.524 & 18.966 & 17.931 & SEGUE\\
  	26 (BHB) & 20.7368 & 38.4931 & 1.0037 & 0.1538 & 0.2964 & 0.1236 & -105.28 & 7.22 & -1.378 & 0.040 & 17.908 & 18.044 & 17.637 & SEGUE\\
  	27 & 20.1517 & 38.6458 & 0.8931 & 0.1369 & 0.2047 & 0.0976 & -101.34 & 5.60 & -2.036 & 0.041 & 17.676 & 18.130 & 17.124 & SEGUE\\
  	\hline
	\end{tabular}
\tablefoot{Identified member stars of Triangulum stream, sorted by $\delta$. Common stars of \citet{2013ApJ...765L..39M} are marked with ``$\star$.'' An outlier identifies in $\mu_{\delta}$ panel of Fig.~\ref{fig:phase-space} is further labeled in ``$\times$.'' Cols.~12-14 are $Gaia$ magnitudes which have been extinction-corrected (see text). The last column indicates which survey the radial velocity and metallicity come from.}
\end{sidewaystable*}

\subsection{NGC 5824 Model Stream}
\label{subsec:mockstream}

\citet{2022ApJ...928...30L} have modeled the disruption of NGC 5824 in a static Milky Way potential plus a moving Large Magellanic Cloud (LMC). As the authors pointed, the model stream matched with observations of Triangulum well. Motivated by this, we also generate our own mock stream to make a similar comparison between the model and data, using the identified member stars above which span a wider sky extent. The model body is nearly identical to that of \citet{2022ApJ...928...30L}, but specific configurations are different, such as the Milky Way potential, the adopted mass and radius of LMC, the velocity dispersion and integration time (see details below). 

We use the Python package \texttt{GALA} \citep{2017JOSS....2..388P}, which is designed for performing common tasks needed in Galactic Dynamics, to model the disruption of NGC 5824. The procedure closely follows that of \citet{2022MNRAS.513..853Y} as applied to NGC 5466. The adopted Milky Way potential consists of a Plummer bulge \citep{1911MNRAS..71..460P}, $\Phi_{\rm bulge}$, two Miyamoto-Nagai disks \citep{1975PASJ...27..533M}, $\Phi_{\rm thin}$ and $\Phi_{\rm thick}$, and a spherical NFW halo \citep{1996ApJ...462..563N}, $\Phi_{\rm halo}$: 

\begin{equation}
  	\Phi_{\rm bulge}(r) = \frac{-GM_{\rm bulge}}{\sqrt{r^2+b_{\rm bulge}^2}}
\end{equation}

\begin{equation}
  	\Phi_{\rm thin/thick}(R,z) = \frac{-GM_{\rm thin/thick}}{\sqrt{R^2+(a_{\rm thin/thick}+\sqrt{z^2+b_{\rm thin/thick}^2})^2}}
\end{equation}

\begin{equation}
  	\Phi_{\rm halo}(r) = \frac{-4\pi G \rho_s r_s^3}{r} {\rm ln}(1+\frac{r}{r_s})
\end{equation}
where $r$ is the Galactocentric radius, $R$ is the cylindrical radius and $z$ is the vertical height. For the bulge and disks, we adopt the parameters from \citet[][Model I]{2017A&A...598A..66P}. The virial mass $M_{\rm virial}$ and concentration $c$ used to initialize the NFW halo are from \citet{2017MNRAS.465...76M}. Those chosen parameters are summarized in Table~\ref{tab:potential}. 

\begin{table}
  	\centering
  	\caption{Adopted parameters for the Galactic potential.}
  	\label{tab:potential}
  	\begin{tabular}{cc}
  		\hline
  		Parameter & Value \\
  		\hline
  		$M_{\rm bulge}$ &  $1.0672\times 10^{10} M_{\sun}$  \\
  		$b_{\rm bulge}$ &  0.3 kpc  \\
  		$M_{\rm thin}$ &  $3.944\times 10^{10} M_{\sun}$  \\
  		$a_{\rm thin}$ &  5.3 kpc  \\
  		$b_{\rm thin}$ &  0.25 kpc  \\
  		$M_{\rm thick}$ &  $3.944\times 10^{10} M_{\sun}$  \\
  		$a_{\rm thick}$ &  2.6 kpc  \\
  		$b_{\rm thick}$ &  0.8 kpc  \\
  		$M_{\rm virial}$ &  $1.37\times 10^{12} M_{\sun}$  \\
  		$c$  &  15.4  \\
  		\hline
  	\end{tabular}
\end{table}

Following \citet{2022MNRAS.510.2437E}, we take a Hernquist Potential \citep{1990ApJ...356..359H} as the internal potential of LMC: 

\begin{equation}
  	\Phi_{\rm LMC}(r') = \frac{-GM_{\rm LMC}}{r'+a_{\rm LMC}}
\end{equation}
where $r'$ is the distance to the LMC center and $M_{\rm LMC}$ and $a_{\rm LMC}$ are set to $10^{11} M_{\sun}$ and 10.2 kpc as well. The position and velocity of LMC are taken from \citet{2018A&A...616A..12G}. 

As for the internal gravity of the GC NGC 5824, we choose a Plummer potential: 

\begin{equation}
  	\Phi_{\rm GC}(r'') = \frac{-GM_{\rm GC}}{\sqrt{r''^2+b_{\rm GC}^2}}
\end{equation}
with a $M_{\rm GC}$ of $7.6\times10^5 M_{\sun}$ and a $b_{\rm GC}$ of 6.51 pc (half-mass radius) \citep{2018MNRAS.478.1520B}. Here $r''$ denotes the distance to the cluster's center. The position and velocity of NGC 5824 come from \citet{2021MNRAS.505.5978V} and \citet[2010 edition]{1996AJ....112.1487H}.

The solar distance to the Galactic center, circular velocity at the Sun and solar velocities relative to the Local Standard of Rest are set to 8 kpc, 220 km\,s$^{-1}$ \citep{2012ApJ...759..131B} and (11.1, 12.24, 7.25) km\,s$^{-1}$ \citep{2010MNRAS.403.1829S}, respectively. In the static Milky Way potential accompanied with a moving LMC, the cluster is initialized 2 Gyr ago\footnote{This integration time is chosen such that the generated mock tidal tail is long enough to completely cover the data.} and integrated forward from then on, releasing two particles (leading and trailing directions respectively) at Lagrange points \citep{2014MNRAS.445.3788G} per 0.05 Myr with a total of 40000 steps. The velocity dispersion is set to 11.9 km\,s$^{-1}$ \citep{2018MNRAS.478.1520B} and the cluster mass is fixed during this process. By doing so, a mock stream for NGC 5824 is obtained as illustrated with the red dots in Fig.~\ref{fig:icrs}. We note that the observed Triangulum (green area) deviates a little from the locus of the mock stream, which also happened in \citet[Fig.~4 therein]{2021ApJ...909L..26B} and \citet[Fig.~8 therein]{2022ApJ...928...30L}. We consider that this deviation between the observation and simulation might be common. 

\subsection{Phase Space}
\label{subsec:comparison}

\begin{figure}
  	\includegraphics[width=\columnwidth]{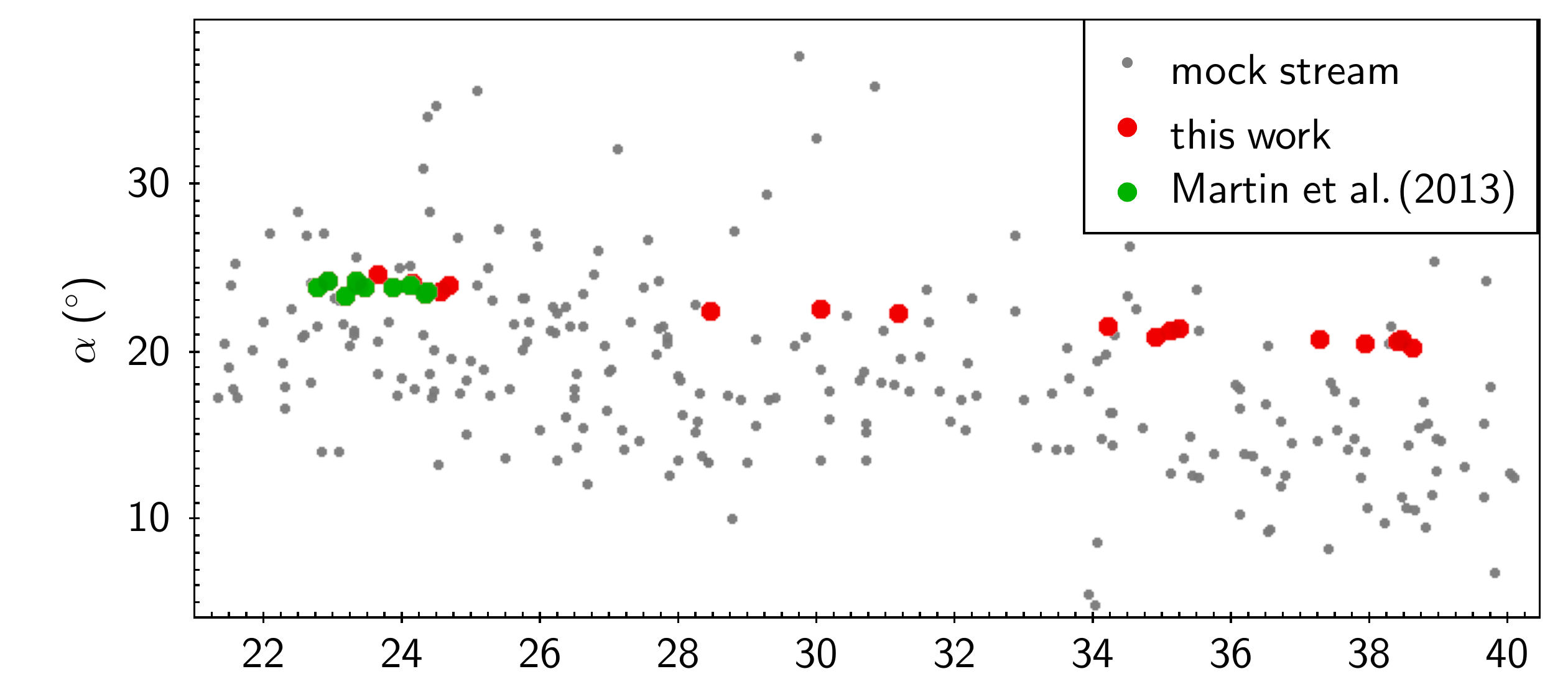}
  	\includegraphics[width=\columnwidth]{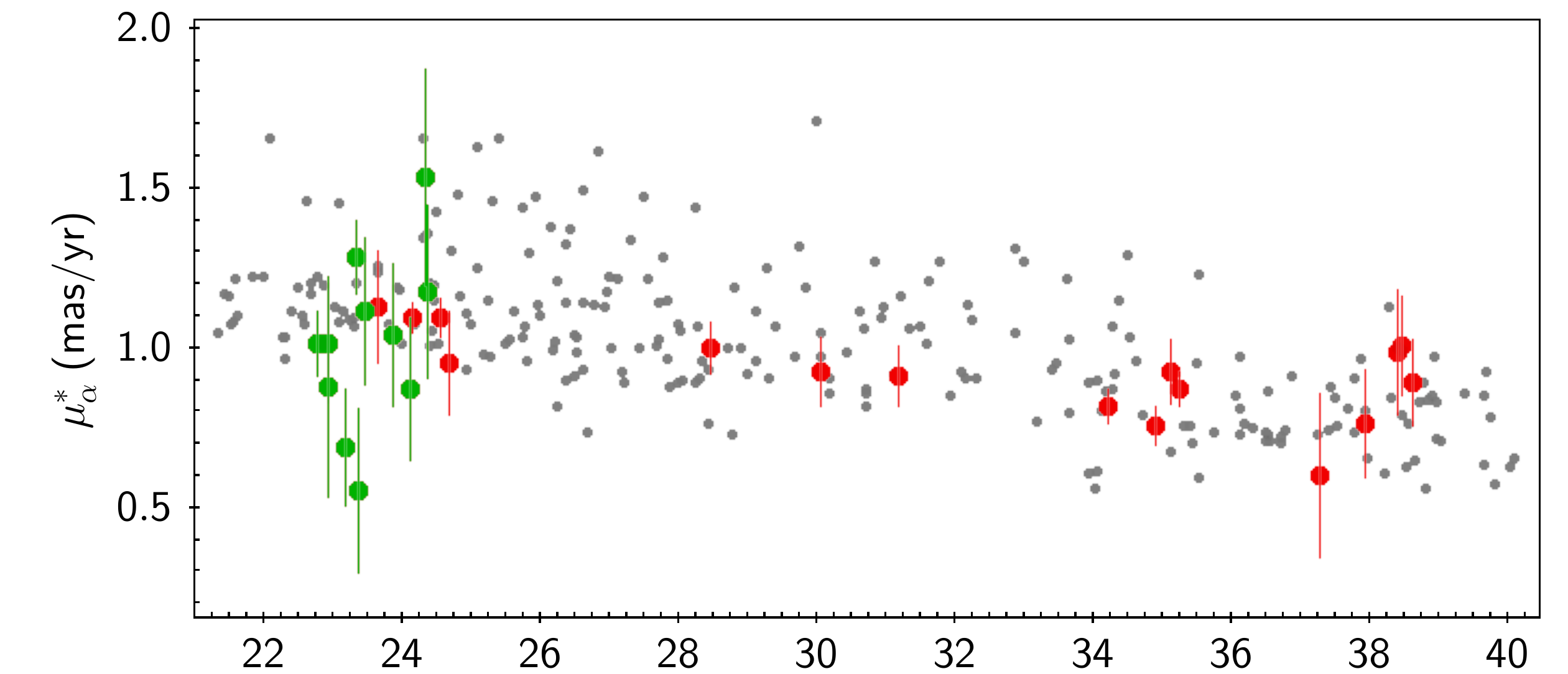}
  	\includegraphics[width=\columnwidth]{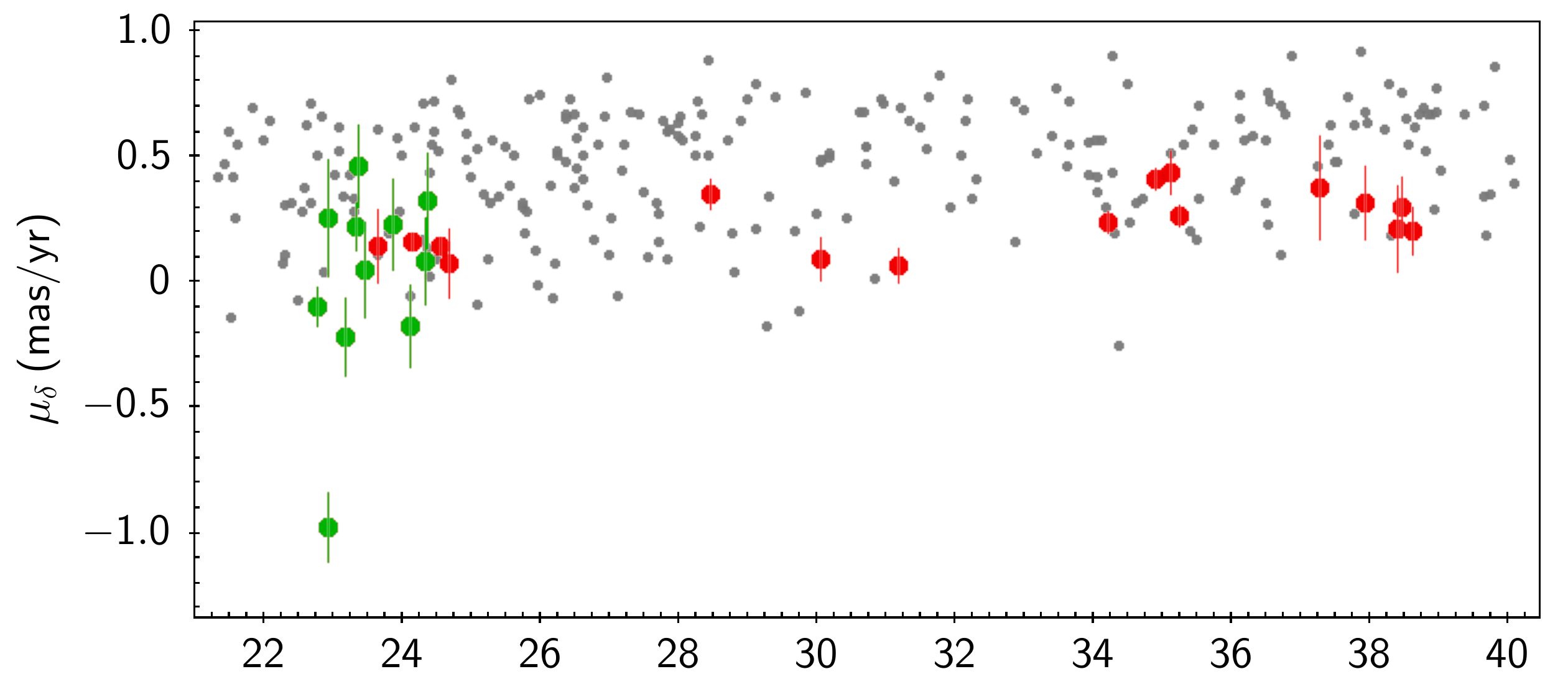}
  	\includegraphics[width=\columnwidth]{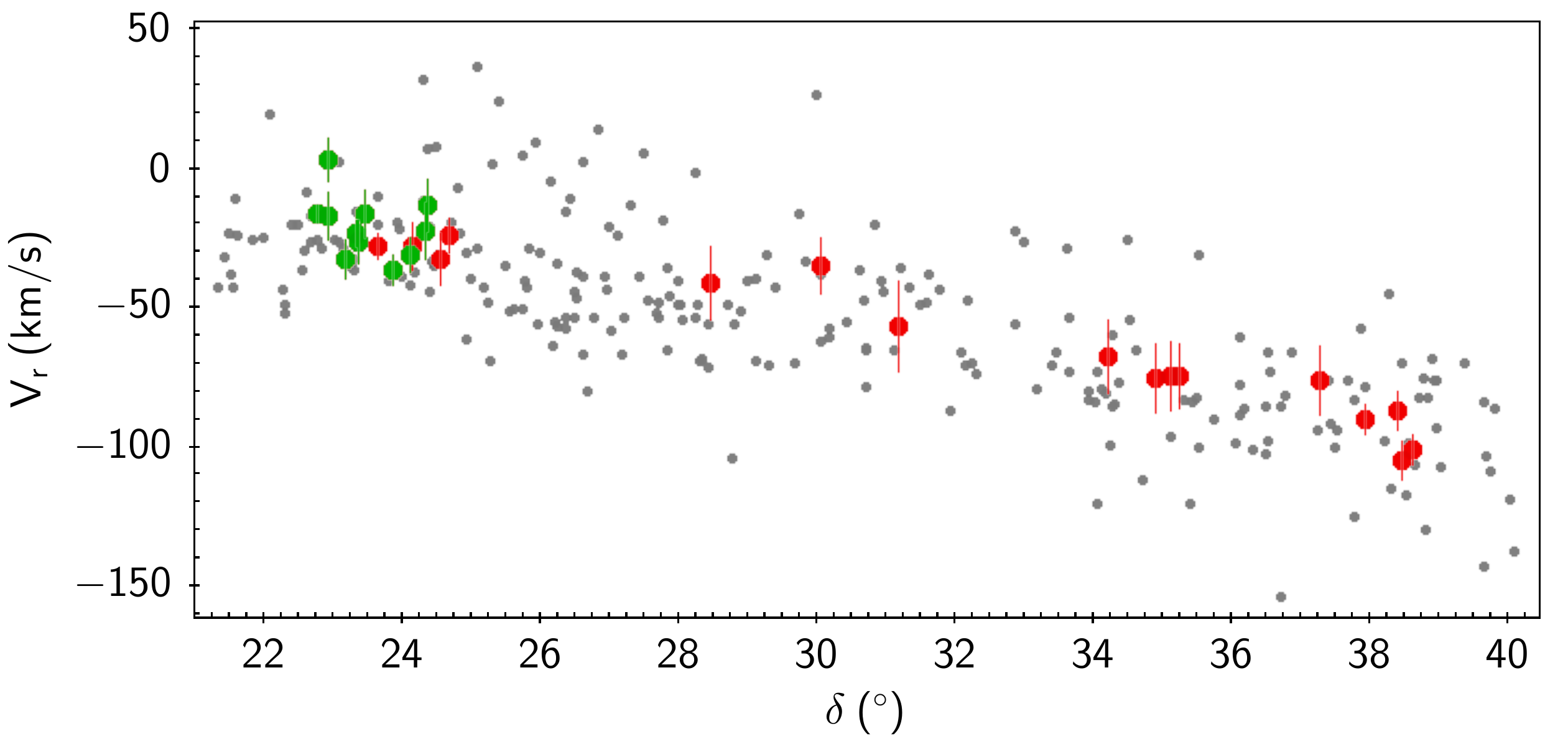}
  	\caption{Right ascension $\alpha$, PMs $\mu^*_{\alpha}$ and $\mu_{\delta}$, and radial velocity $V_r$ as a function of declination $\delta$ are presented from top to bottom. The gray dots represent the stream particles within the same sky area as Triangulum. The green and red points represent the stream member stars.}
  	\label{fig:phase-space}
\end{figure}

We compare the Triangulum member stars to the model stream of NGC 5824 in phase space. In Fig.~\ref{fig:phase-space}, right ascension $\alpha$, PMs $\mu^*_{\alpha}$ and $\mu_{\delta}$, and radial velocity $V_r$ as a function of declination $\delta$ are presented from top to bottom. The gray dots represent the stream particles within the same sky area as Triangulum. The member stars are shown in the red and green points.

It can be seen that even though the selection process of member stars in Sect.~\ref{subsec:members} is completely independent of the model, the stream particles show good consistency with the observations in phase space. We note an outlier that falls too far from the others in $\mu_{\delta}$ plane. This star was selected by \citet{2013ApJ...765L..39M} based on sky position, radial velocity, metallicity and CMD, when PM measurements were unavailable. We mark it with ``$\times$'' in Table~\ref{tab:members} and remove it in subsequent analysis. Furthermore, we do not show distance plane here because there is some confusion, and we present a discussion about it in Sect.~\ref{sec:discussion}.

\subsection{Metallicity and CMD}

To further examine whether Triangulum is stripped from GC NGC 5824, we compare them on the basis of metallicity and CMD. 

\begin{figure}
  	\includegraphics[width=\columnwidth]{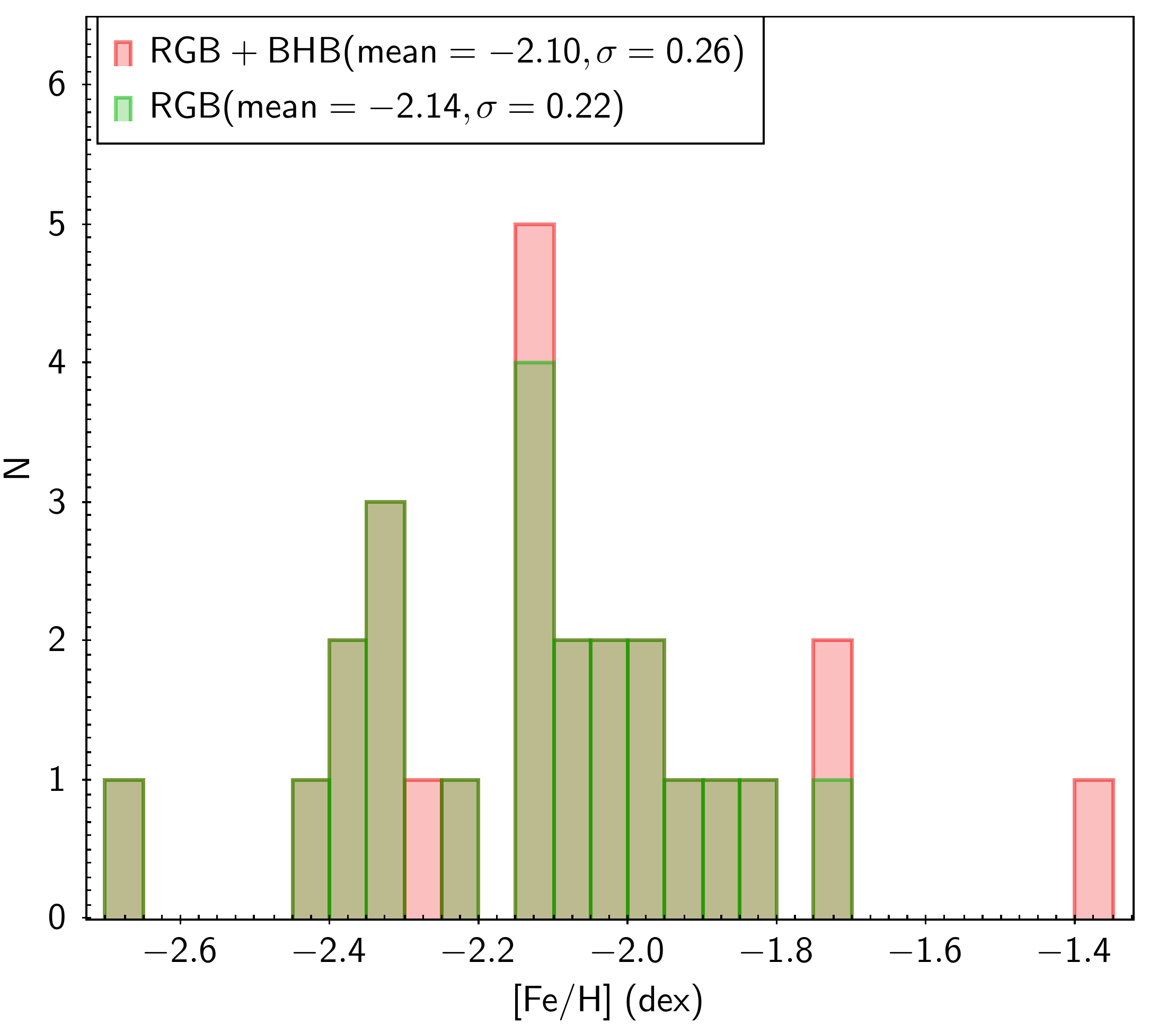}
  	\caption{The metallicity distribution of Triangulum member stars. The red bars represent the whole sample and the green bars correspond to only RGB stars. }
  	\label{fig:hist}
\end{figure}

The metallicity distribution of Triangulum members is presented in Fig.~\ref{fig:hist}. There are 4 blue horizontal branch (BHB) stars and 22 red giant branch (RGB) stars. For the whole sample, the mean value $\langle \rm [Fe/H] \rangle$ = -2.10 and standard deviation $\sigma_{\rm [Fe/H]}$ = 0.26 dex are consistent with those of \citet{2013ApJ...765L..39M} ($\langle \rm [Fe/H] \rangle$ = -2.2, $\sigma_{\rm [Fe/H]}$ = 0.3 dex). Picking out RGB stars separately is aimed for a comparison to some chemical researches on NGC 5824. \citet{2018ApJ...859...75M} analyzed 87 RGB stars of the cluster and obtained a metallicity distribution peaked at [Fe/H] = -2.11 dex, which is very similar to $\langle \rm [Fe/H] \rangle$ = -2.14 dex here. The observed scatter $\sigma_{\rm [Fe/H]}$ = 0.22 dex could probably be caused by observational uncertainties in low-resolution spectra ($R\sim$1800).

To compare Triangulum with GC NGC 5824 in CMD, we need to know the stream's distance. \citet{2011ApJ...738...79X} estimated distances of $\sim$ 5000 BHB stars by matching them in ($u - g$, $g - r$) space to theoretical colors for BHB stars with a series of absolute magnitudes. The individual distances of 4 BHB stars in our sample can be obtained from this catalog: 28.8, 26.9, 30.6 and 26.0 kpc for stars with No. 4, 8, 24 and 26 in Table~\ref{tab:members}, respectively. This yields a median distance of 27.85 kpc, close to 26 kpc proposed by \citet{2012ApJ...760L...6B}. In addition, we also estimate distances to all 26 stars (see Fig.~\ref{fig:dist}) using the method from \citet{2015AJ....150....4C}, which is a Bayesian approach with likelihood estimated via comparison of spectroscopically derived atmospheric parameters to a grid of stellar isochrones, and returns a posterior probability density function for star's absolute magnitude. This yields a median value at 33 kpc similar to 35 kpc estimated by \citet{2013ApJ...765L..39M}. We adopt the distance to Triangulum stream as $\sim$ 30 kpc, which is a median value between BHB distance and our estimate. 

\begin{figure}
  	\includegraphics[width=\columnwidth]{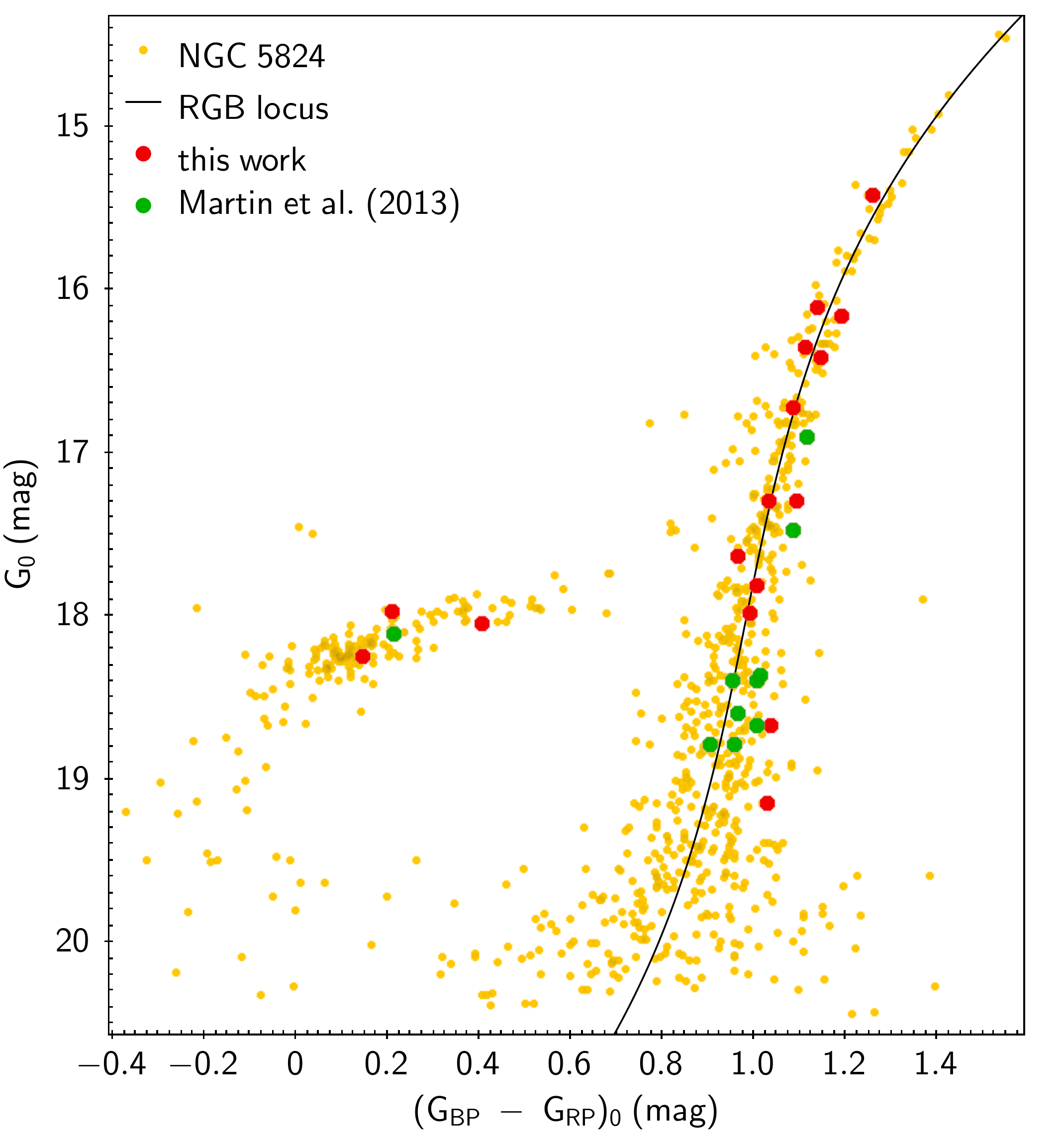}
  	\caption{The orange dots represent GC NGC 5824 stars. The red and green dots represent Triangulum members.  The black line denotes the RGB locus obtained by directly fitting the RGB stars with a third-order polynomial.}
  	\label{fig:cmd_done}
\end{figure}

In CMD, we move the member stars from 30 to 32.1 kpc, where GC NGC 5824 is located \citep[2010 edition]{1996AJ....112.1487H}, and find that they match well as shown in Fig.~\ref{fig:cmd_done}. The cluster stars here marked in the orange dots are obtained through sky and PM selections as instructed by \citet{2021A&A...645A.116K}. Specifically, we retrieve stars within the tidal radius $r_t$ = 5.73$'$ of NGC 5824 \citep[2010 edition]{1996AJ....112.1487H} and clean the data following procedures as described in Sect.~\ref{sec:data}. A 2D Gaussian mixture model consisting of two Gaussians is then fitted in PM space to decompose the cluster and field stars apart. For the cluster component, we get the center ($\mu^*_{\alpha}$, $\mu_{\delta}$) = (-1.193, -2.235) with the intrinsic dispersion ($\sigma^{\rm in}_{\mu^*_{\alpha}}$, $\sigma^{\rm in}_{\mu_{\delta}}$) = (0.424, 0.360) mas yr$^{-1}$, where the center is very close to (-1.189, -2.234) mas yr$^{-1}$ measured by \citet{2021MNRAS.505.5978V}. The cluster stars are selected as those whose PMs, within uncertainties, match the PMs and dispersion of NGC 5824: $\{\mu^*_{\alpha}\pm \sigma_{\mu^*_{\alpha}}, \mu_{\delta}\pm \sigma_{\mu_{\delta}}\}^{\rm star}$ $\leq$ $\{\mu^*_{\alpha}\pm \sigma^{\rm in}_{\mu^*_{\alpha}}, \mu_{\delta}\pm \sigma^{\rm in}_{\mu_{\delta}}\}^{\rm cluster}$. The black line denotes the RGB locus obtained by fitting the RGB stars directly with a third-order polynomial, which is used in Sect.~\ref{subsec:mf} to assign weights in CMD.  

The connection between the stream and the cluster, based on the three aspects above, confirms that Triangulum was disrupted from GC NGC 5824. In other words, the stream can be treated as a part of the cluster's leading tail.

\section{Detecting the Trailing Tail}
\label{sec:match-filter}

Motivated by the existence of leading tail for NGC 5824, in this section we aim to search for its trailing tail.

\subsection{A Modified Matched Filter Method}
\label{subsec:mf}

Combining PMs and CMD together to search for extra-tidal structures of GCs has proved to be an effective way \citep[e.g.,][]{2019MNRAS.489.4565K,2019MNRAS.483.1737K,2021A&A...645A.116K}. Here we adopt the method from \citet{2019ApJ...884..174G} who applied a modified matched filter technique and successfully detected a 50$\degr$ long tidal tail for GC M5. 

Stars fetched in Sect.~\ref{sec:data} are assigned weights based on their locations in CMD and PM space. In CMD, individual stars in the NGC 5824 field are assigned weights according to their color differences from the cluster locus, assuming a Gaussian error distribution:

\begin{equation}
	w_{\rm CMD} = \frac{1}{\sqrt{2\pi} \sigma_{color}} {\rm exp}
	\left[ -\frac{1}{2} \left(\frac{color - color_0}{\sigma_{color}} \right)^2  \right]  .
\end{equation}
Here $color$ and $\sigma_{color}$ denote $G_{BP}$ - $G_{RP}$ and corresponding errors. Color errors are simply calculated through $\sqrt{\sigma^2_{G_{BP}} + \sigma^2_{G_{RP}}}$ where $\sigma_{G_{BP}}$ and $\sigma_{G_{RP}}$ are obtained with a propagation of flux errors (see CDS website \footnote{\url{https://vizier.u-strasbg.fr/viz-bin/VizieR-n?-source=METAnot&catid=1350&notid=63&-out=text}.}). $color_0$ is determined by the cluster RGB locus (the black line in Fig.~\ref{fig:cmd_done}) at a given $G$ magnitude of a star. During assigning weights, we do not include $\sigma_{G}$ since uncertainties in $G$ band are much smaller than those in $G_{BP}$ and $G_{RP}$ (on the order of $\sim$ 0.1) for $Gaia$ photometry. Stars from $G$ = 15 mag (tip of the cluster's RGB) to the $Gaia$ limit $G \simeq$ 21 mag are investigated. 

\begin{figure*}
	\centering
	\includegraphics[width=0.8\linewidth]{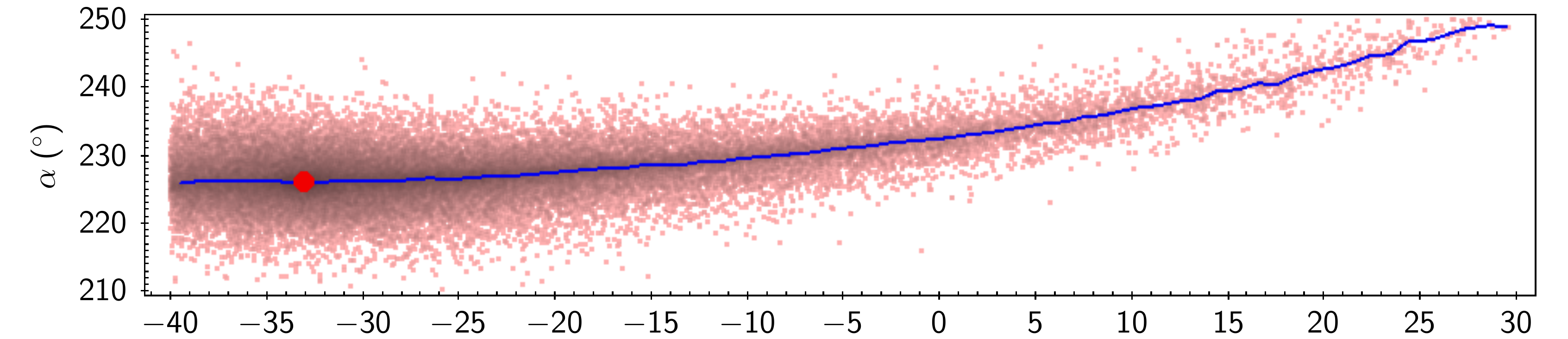}
	\includegraphics[width=0.8\linewidth]{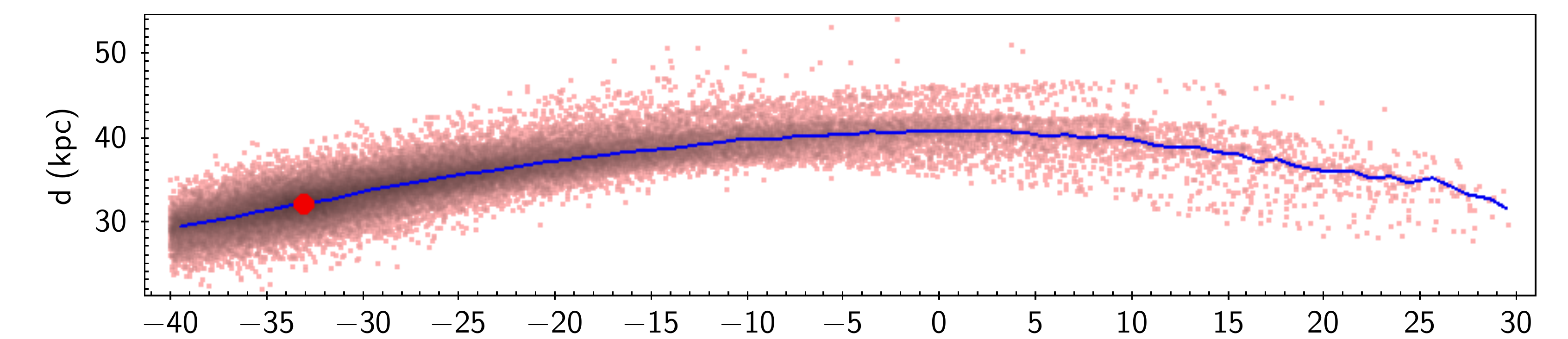}
	\includegraphics[width=0.8\linewidth]{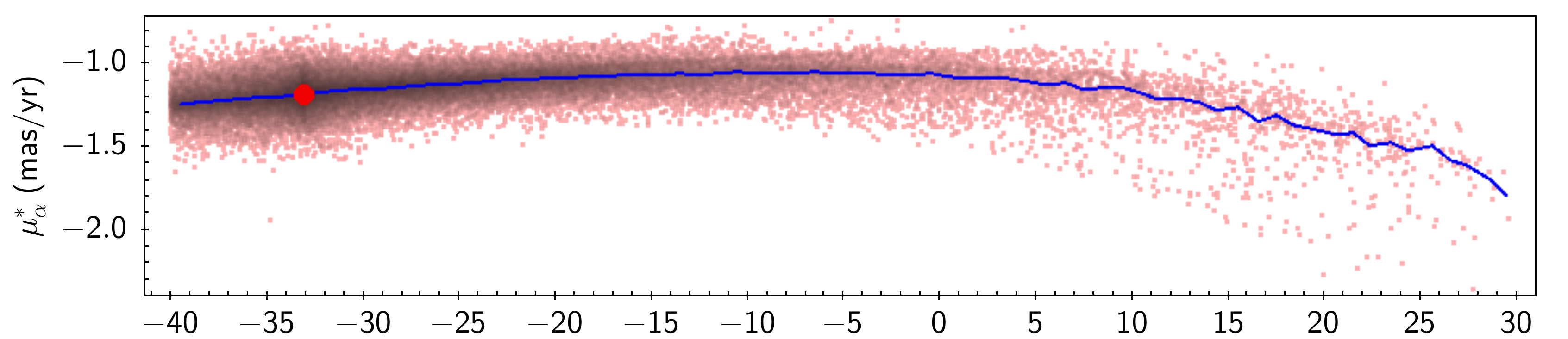}
	\includegraphics[width=0.8\linewidth]{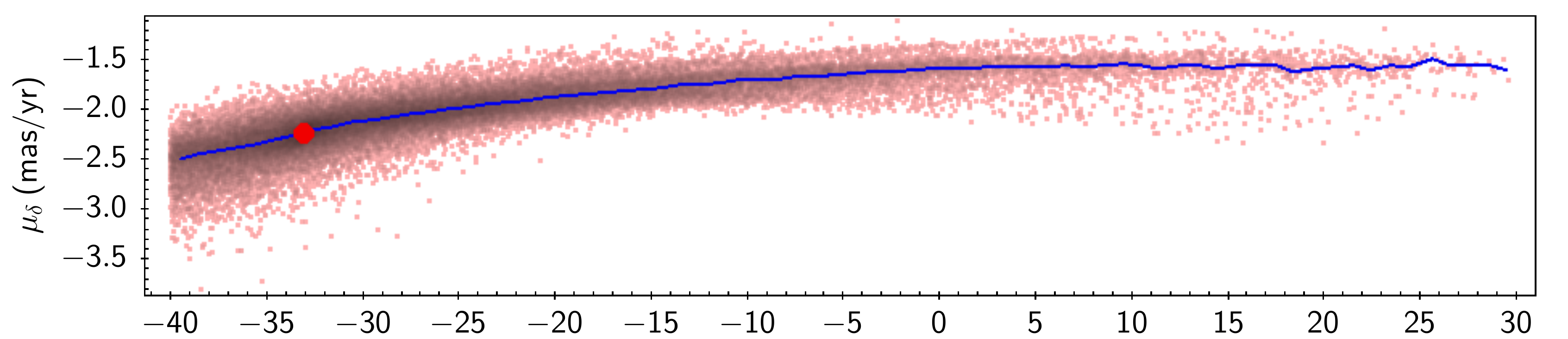}
	\includegraphics[width=0.8\linewidth]{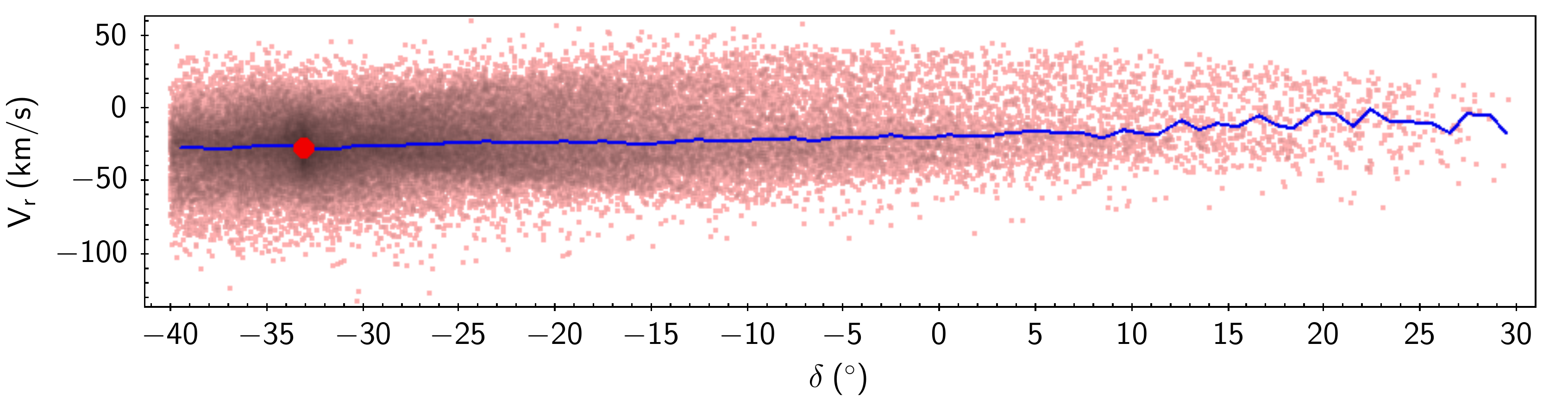}
	\caption{The planes of $\alpha$, heliocentric distance, proper motion in $\alpha$ and $\delta$, and radial velocity as a function of $\delta$, are shown from the top to the bottom, respectively. The pink dots represent the model stream particles within the NGC 5824 field. The red circle represents GC NGC 5824. The blue lines denote medians of y-axis values in each $\delta$ bin with a bin width of 1$\degr$. }
	\label{fig:locus}
\end{figure*}

\begin{figure}
	\includegraphics[width=\columnwidth]{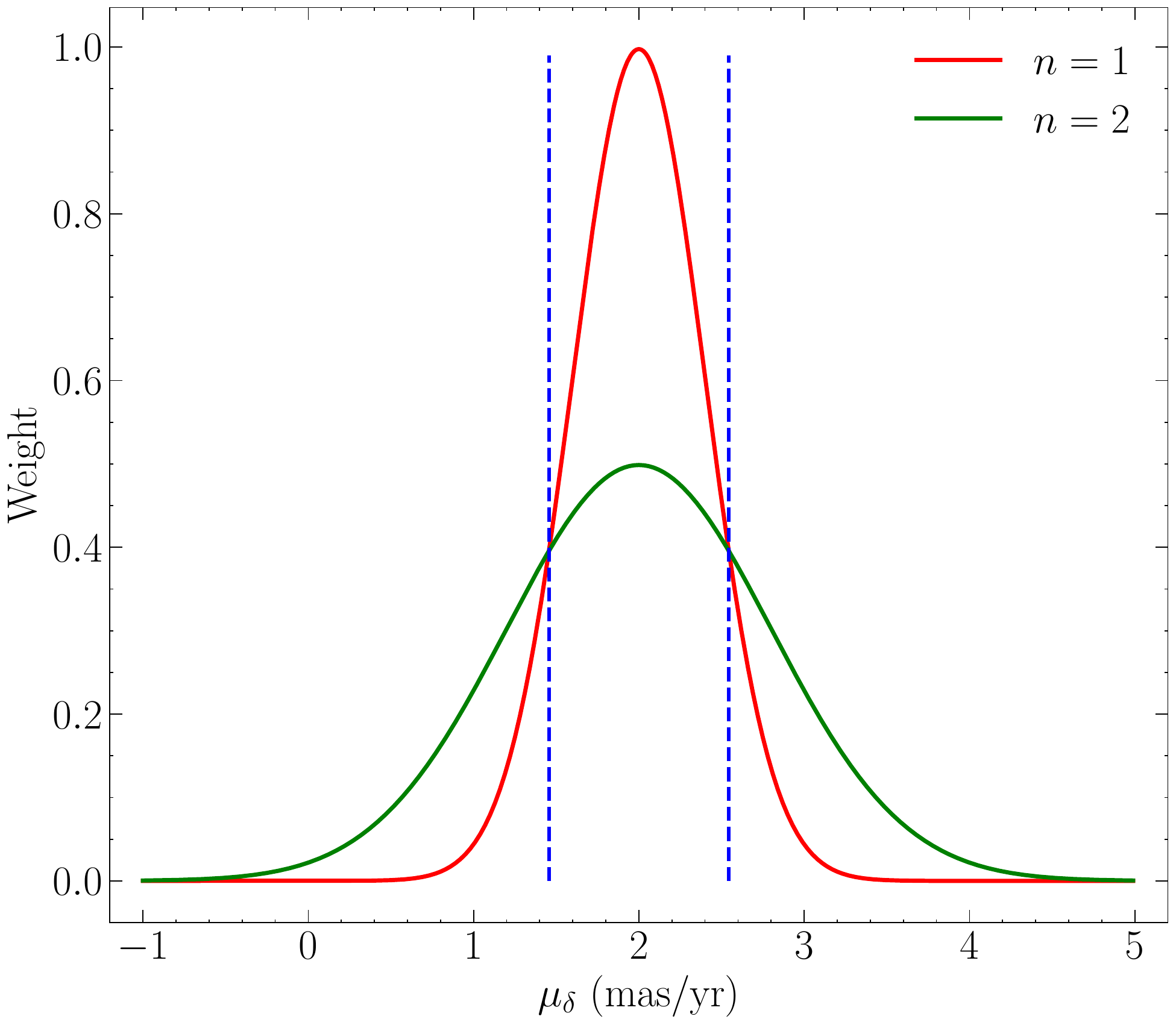}
	\caption{Illustration for using $n$ = 2 in Eq.~(\ref{eq7}). The red and green lines represent Gaussians centered at 2 with sigma = 0.4 and 0.8 mas yr$^{-1}$, respectively. }
	\label{fig:n1_vs_n2}
\end{figure}

The PMs of the model stream generated in Sect.~\ref{subsec:mockstream} are further employed to weight stars. Fig.~\ref{fig:locus} shows the stream particles within the NGC 5824 field in phase space, which serves as an estimate to the real stream. In PM space, weights are computed as:

\begin{equation}
	w_{\rm PMs} = \frac{1}{2\pi n^2 \sigma_{\mu^*_{\alpha}}\sigma_{\mu_{\delta}}} {\rm exp}
	\left\lbrace -\frac{1}{2} \left[ \left( \frac{\mu^*_{\alpha} - \mu^*_{\alpha,0}}{n\sigma_{\mu^*_{\alpha}}} \right)^2 +
	\left( \frac{\mu_{\delta} - \mu_{\delta,0}}{n\sigma_{\mu_{\delta}}} \right)^2 \right] \right\rbrace .
	\label{eq7}
\end{equation}
$\mu^*_{\alpha}$, $\mu_{\delta}$, $\sigma_{\mu^*_{\alpha}}$ and $\sigma_{\mu_{\delta}}$ are measured PMs and corresponding errors. $\mu^*_{\alpha,0}$ and $\mu_{\delta,0}$ are the components of PMs predicted at
each star's $\delta$ based on the model stream's locus (blue lines of PM panels in Fig.~\ref{fig:locus}). The locus is obtained by dividing the particles into $\delta$ bins (bin width = 1$\degr$) and calculating medians of PMs in each bin. It is worth noting that PM errors are multiplied by $n$ and we choose a moderate $n$ = 2 here, which is designed to allow some deviations between the model and observations. This can be illustrated using a one-dimensional example (see Fig.~\ref{fig:n1_vs_n2}). Assume that we are going to assign a weight to a stream star (if exist) with $\mu_{\delta}$ = $x$ and $\sigma_{\mu_{\delta}}$ = 0.4 mas yr$^{-1}$. The $\mu_{\delta,0}$ predicted by the model stream at the star's $\delta$ is 2 mas yr$^{-1}$. The star's weight will be determined by a Gaussian with mean = 2 and sigma = 0.4 ($n$ = 1, red line) or 0.8 ($n$ = 2, green line). If the model predicts the stream very well, that is $x$ is very close to 2 mas yr$^{-1}$, the red line ($n$ = 1) will give a higher weight to the star apparently. However, the model stream is just an approximation to the real one and it is likely that there are small deviations between them, which might lead to that $x$ falls out of the blue dashed lines. When this happens, the green line ($n$ = 2) gives a higher weight. We have compared results using different $n$ values and verified that $n$ = 2 is the most favorable. 

Finally, stars weights are obtained by multiplying $w_{\rm CMD}$ and $w_{\rm PMs}$, and then summed in $0.2\degr \times 0.2\degr$ sky pixels to expose structures.

\subsection{Results}

\begin{figure*}
	\centering
	\includegraphics[width=\linewidth]{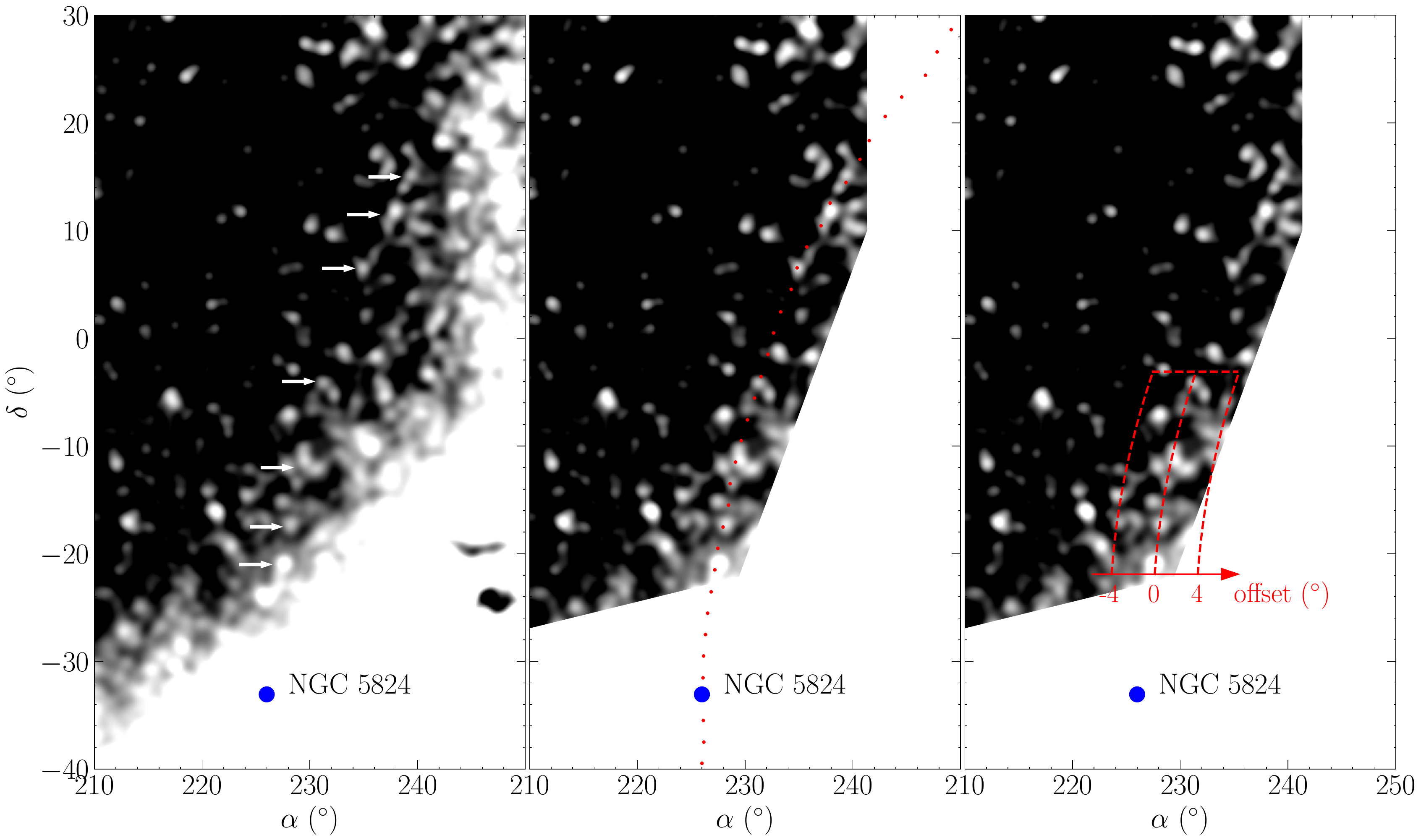}
	\caption{Log stretch of a matched filter map in the NGC 5824 field. The sky pixel width is 0.2$\degr$ and the map is smoothed with a Gaussian kernel of $\sigma$ = 0.5$\degr$. Three panels present the same map. The white arrows in left panel point the stream features. The locus of the model stream is overplotted in the small red dots in middle panel. The right panel illustrates the way of creating the stream's lateral profile (see text). Bottom-right region close to the Galactic disk is masked in the middle and right panels. }
	\label{fig:tail_radec}
\end{figure*}

A weighted sky map is obtained after applying the above method to data in the cluster field and shown in the left panel of Fig.~\ref{fig:tail_radec}. To make the stream look more prominent, pixels with summed weights $>$ 80 and $<$ 2 are masked such that too strong noises and weak background are not shown. The map is then smoothed with a Gaussian kernel of $\sigma$ = 0.5$\degr$. The stretch is logarithmic, with brighter areas corresponding to higher weight regions. The blue circle on the bottom marks the location of NGC 5824. The white bottom-right corner is due to being close to the Galactic disk, which is further masked in the middle and right panels. 

Due to the photometric depth of $Gaia$, the cluster's main sequence stars are not observable and only RGB stars can be used to trace the underlying trailing tail, which are much fewer than the former. However, some stream-like signals are still detected. In the left panel, it is clear that there are several structures (marked with arrows) with higher weights between $\delta \simeq$ -21 $-$ -4$\degr$ that could be connected smoothly and likely extended from NGC 5824. In the middle panel, we overplot the trajectory of the model stream (small red dots) and find that it passes through the structures well. An additional segment of $\delta \simeq$ 6 $-$ 16$\degr$ is a farther extension of the stream. There is a gap in the middle at $\delta \simeq$ -4 $-$ 6$\degr$ corresponding to the most distant range of the model stream (see the distance panel in Fig.~\ref{fig:locus}), where many RGB stars might have been darker than 21 mag. The detected signature traces the cluster's trailing tail to $\sim 50\degr$ whose path can be roughly fitted using

\begin{equation}
	\alpha = 4.07\times 10^{-5}\delta^3 + 6.68\times 10^{-3}\delta^2 + 0.37\delta +232.45
\end{equation}
where -33$\degr$ $<$ $\delta$ $<$ 16$\degr$.

\begin{figure}
	\includegraphics[width=\columnwidth]{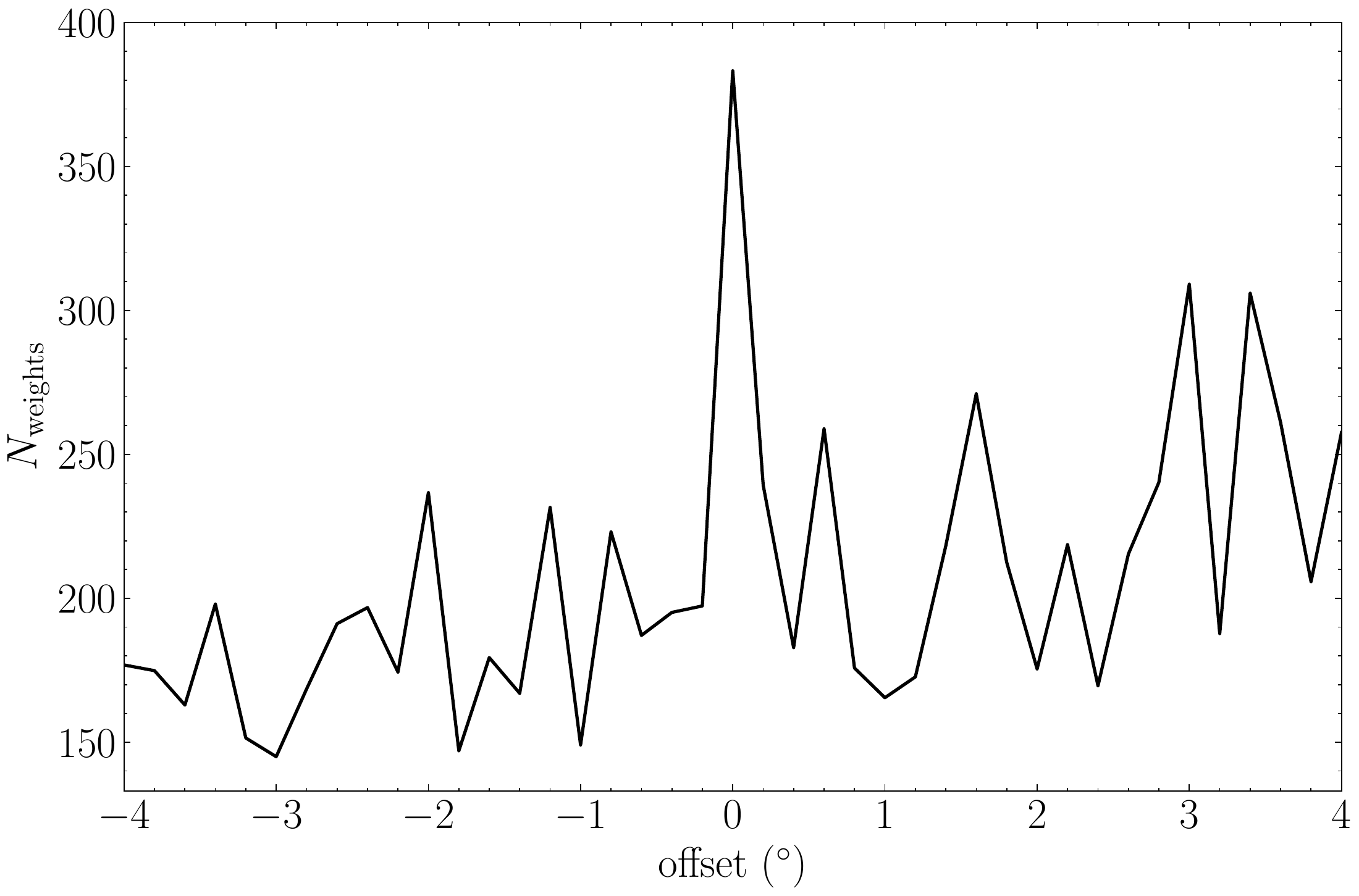}
	\caption{The stream one-dimensional profile. The offset coordinate is defined as deviation from the stream along $\alpha$ direction (see Eq.~(\ref{eq9})).}
	\label{fig:profile}
\end{figure}

In the right panel of Fig.~\ref{fig:tail_radec}, stars enclosed by the red lines are selected to calculate the statistical significance of the stream. The $\delta$ range is $-22\degr - -3\degr$. The central dashed line represents a more precise description to the stream of this region, which is given by
\begin{equation}
	\alpha = 7.15\times 10^{-3}\delta^2 + 0.38\delta + 232.58 + {\rm offset} 
	\label{eq9}
\end{equation}
with offset = 0$\degr$. The left and right boundaries correspond to offset = -4 and 4$\degr$, respectively. A bin width = 0.2$\degr$ is used and at offset = -4, -3.8, -3.6..., weights of stars around Eq.~(\ref{eq9}) $\pm 0.1\degr$ are integrated to create a lateral profile of the stream as displayed in Fig.~\ref{fig:profile}. The central peak at offset = 0$\degr$ represents the stream feature. The larger random counts at positive side are caused by higher stellar density near the disk. The significance is defined as $S = (w_{stream} - w_{background}) / \sigma_{background}$, where $w_{stream}$ is the stream signal and $w_{background}$ and $\sigma_{background}$ are the mean and standard deviation of weights for off-stream regions 0.5$\degr$ $<$ |offset| $<$ 4$\degr$. We get S = 7.5 and 3.6 for negative and positive sides, respectively, and S is 4.3 if both are considered. It can be inferred from Fig.~\ref{fig:profile} that the stream's width is expected to be $\lesssim 0.2\degr$ because signals drop back to the level of background when |offset| $>$ 0.1$\degr$ which means that there are few stream signals beyond this range. If we adopt $d$ = 39 kpc for this segment based on the model, the physical width is $\lesssim$ 136 pc. 

\subsection{A Part of Cetus?}

\citet{2021ApJ...909L..26B} pointed that GC NGC 5824 and Cetus \citep{2009ApJ...700L..61N}, which is a stellar stream with a dwarf galaxy origin, have very close orbital energies and angular momenta. Similar orbital trajectories between them are also demonstrated in \citet{2020ApJ...905..100C}. This arises a question: do those features on the trailing side of the cluster belong to Cetus stream?

Combining the results here with previous researches on Cetus, we present 4 reasons for that the detected features are indeed related to the trailing tail of NGC 5824. 
\begin{enumerate}
\item The width of features in Fig.~\ref{fig:tail_radec} is only $\lesssim 0.2\degr$, which is thin compared to a stream produced by a dwarf galaxy.
\item Cetus stars should have a relatively spread distribution in CMD. However, the stream features indicated with arrows in Fig.~\ref{fig:tail_radec} disappear if the RGB locus used to weight stars in CMD is shifted either blueward or redward by 0.1 mag, which means that they are exactly corresponding to NGC 5824. 
\item \citet{2020ApJ...905..100C} pointed that GC NGC 5824 should not be the core of Cetus, implying that there is no direct connection between the cluster and Cetus stream. Furthermore, \citet{2021arXiv211205775Y} concentrated on searching for Cetus's members using data covering the cluster but they did not detect any densely populated structure around NGC 5824. Hence the features should not be a part of Cetus.
\item Triangulum as a piece of the leading tail also provides a weak evidence of existence for the trailing tail. 

\end{enumerate}

\section{Discussion}
\label{sec:discussion}

\begin{figure}
	\includegraphics[width=\columnwidth]{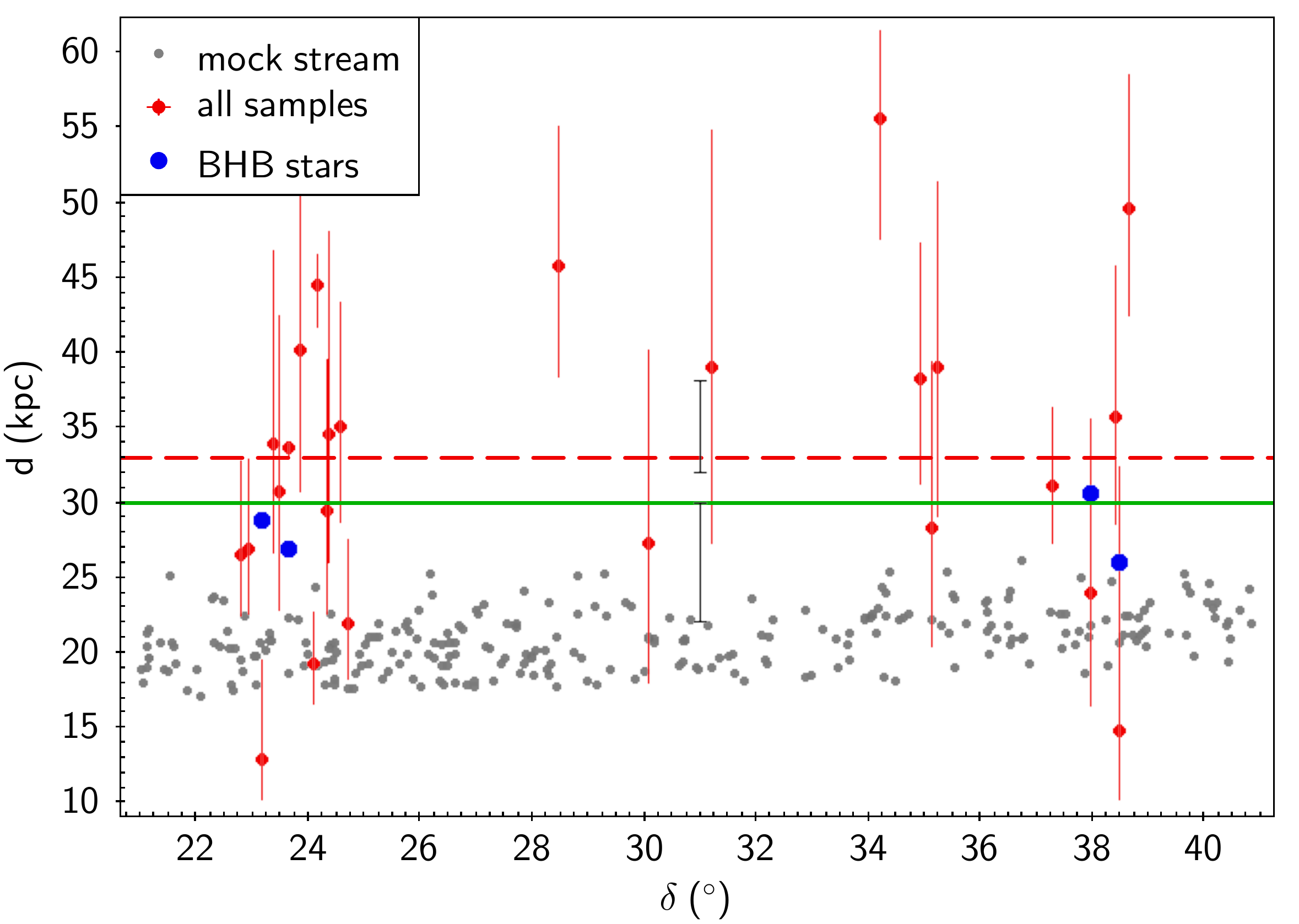}
	\caption{Heliocentric distance as a function of $\delta$. The gray dots represent the stream particles. The blue points represent 4 BHB stars in our samples, whose distances come from \citet{2011ApJ...738...79X}. The red dots and error bars present distances and corresponding errors for all member stars estimated using the method of \citet{2015AJ....150....4C}. The red dashed line corresponds to their median value 33 kpc. The green solid line marks adopted distance 30 kpc. Two black error bars represent 26 $\pm$ 4 kpc \citep{2012ApJ...760L...6B} and 35 $\pm$ 3 kpc \citep{2013ApJ...765L..39M}. }
	\label{fig:dist}
\end{figure}

During comparing Triangulum to the model stream in distance, we find some incompatibility and show them in Fig.~\ref{fig:dist}. \citet{2012ApJ...760L...6B} estimated a Triangulum's distance of 26 $\pm$ 4 kpc (the lower black error bar) while \citet{2013ApJ...765L..39M} proposed 35 $\pm$ 3 kpc (the upper black error bar) for the stream. As mentioned above, we adopt a distance of 30 kpc (the green solid line) and find that the member stars match with GC NGC 5824 well in CMD. However, under a static Milky Way potential plus a moving LMC, the resulting model stream predicts that Triangulum's distance should be about 20 - 25 kpc (gray dots), which is true in both this work and \citet{2022ApJ...928...30L} (the second panel of Fig.~8 therein). This arises a confusion: why is there such a difference? 

\citet{2014ApJ...793...62S} presented an analysis on TriAnd1 ($d \sim$ 20 kpc) and TriAnd2 ($d \sim$ 28 kpc) \citep{2007ApJ...668L.123M}, other two stellar substructures in the direction of M31 and M33. They show that even though the two structures are separated by more than 5 kpc in distance, they are indistinguishable in radial velocity and PMs. We note that this kinematic feature is very similar to that of Triangulum when compared to the model stream. The real and mock streams are separated by more than 5 kpc as well but their trends in phase space are still in concordance. Considering that the stream and those structures are exactly in the same region, it is very likely that Triangulum has been affected by the mechanism which leads to TriAnd1 and TriAnd2. Specifically, either related to a dwarf galaxy \citep{2014ApJ...793...62S} or the Galactic disk \citep{2015ApJ...801..105X}, some process that created TriAnd1 and TriAnd2 might push Triangulum farther away (30 kpc) from where it should be (20 - 25 kpc). We anticipate that this prediction could be proved by later simulations on the formation of TriAnd overdensities.

It is also worth nothing that there is another stream segment named Turbio \citep{2018ApJ...862..114S} between Triangulum and GC NGC 5824 that was considered to be disrupted from the cluster based on their similar dynamics in \citet{2021ApJ...909L..26B} and \citet{2022ApJ...928...30L}. We do not inspect this stream due to lack of spectroscopic data. It is expected that upcoming observations will be able to provide more details on connections between Turbio and the cluster, and even more opportunities of searching for other stream segments on the leading side. If these can be confirmed, NGC 5824 tidal tails would be the longest cold stream ever discovered in the Milky Way.

\section{Conclusions}
\label{sec:summary}

We first validate the connection between Triangulum stream and NGC 5824. A total of 26 stream member stars are selected and 16 of them are newly identified. We model the cluster's disruption under a static Milky Way potential accompanied with a moving LMC. The real stream is compared to the mock one in phase space and consistent trends can be found. In metallicity and CMD, the member stars and the cluster are also in good agreement. These results support the previous statement that Triangulum originates from GC NGC 5824 \citep{2021ApJ...909L..26B,2022ApJ...928...30L}. 

Given that Triangulum can be considered as a segment of the cluster's leading tail, we examine the existence of its trailing tail. Using a matched-filer method that combines CMD and PMs to weight stars, we find a $\sim$ 50$\degr$ trailing tail for GC NGC 5824. The features match with the model stream well. Although the signals are tenuous and discrete, a peak of $>$ 3$\sigma$ over background noises can be still discerned in the lateral stream profile, from which we estimate that its width is $\lesssim$ 0.2$\degr$. We expect that follow-up observations will be able to provide more details about the NGC 5824 stream.

\begin{acknowledgements}
	
We thank the anonymous referee, whose comments greatly improved this publication. This study was supported by the National Natural Science Foundation of China under grant nos 11988101, 11973048, 11927804, 11890694 and 11873052, and the National Key R\&D Program of China, grant no. 2019YFA0405500. This work (MNI) is also supported by JSPS KAKENHI Grant Number 20H05855 and the GHfund A (202202018107). We acknowledge the support from the 2m Chinese Space Station Telescope project: CMS-CSST-2021-B05.

Guoshoujing Telescope (the Large Sky Area Multi-Object Fiber Spectroscopic Telescope LAMOST) is a National Major Scientific Project built by the Chinese Academy of Sciences. Funding for the project has been provided by the National Development and Reform Commission. LAMOST is operated and managed by the National Astronomical Observatories, Chinese Academy of Sciences.

This work presents results from the European Space Agency (ESA) space mission $Gaia$. $Gaia$ data are being processed by the $Gaia$ Data Processing and Analysis Consortium (DPAC). Funding for the DPAC is provided by national institutions, in particular the institutions participating in the $Gaia$ MultiLateral Agreement (MLA). The $Gaia$ mission website is \url{https://www.cosmos.esa.int/gaia}. The $Gaia$ archive website is \url{https://archives.esac.esa.int/gaia}.

Funding for the Sloan Digital Sky Survey IV has been provided by the Alfred P. Sloan Foundation, the U.S. Department of Energy Office of Science, and the Participating Institutions. 

SDSS-IV acknowledges support and resources from the Center for High Performance Computing  at the University of Utah. The SDSS website is \url{www.sdss.org}.

SDSS-IV is managed by the Astrophysical Research Consortium for the Participating Institutions of the SDSS Collaboration including the Brazilian Participation Group, the Carnegie Institution for Science, Carnegie Mellon University, Center for Astrophysics | Harvard \& Smithsonian, the Chilean Participation Group, the French Participation Group, Instituto de Astrof\'isica de Canarias, The Johns Hopkins University, Kavli Institute for the Physics and Mathematics of the Universe (IPMU) / University of Tokyo, the Korean Participation Group, Lawrence Berkeley National Laboratory, Leibniz Institut f\"ur Astrophysik Potsdam (AIP),  Max-Planck-Institut f\"ur Astronomie (MPIA Heidelberg), Max-Planck-Institut f\"ur Astrophysik (MPA Garching), Max-Planck-Institut f\"ur Extraterrestrische Physik (MPE), National Astronomical Observatories of China, New Mexico State University, New York University, University of Notre Dame, Observat\'ario Nacional / MCTI, The Ohio State University, Pennsylvania State University, Shanghai Astronomical Observatory, United Kingdom Participation Group, Universidad Nacional Aut\'onoma de M\'exico, University of Arizona, University of Colorado Boulder, University of Oxford, University of Portsmouth, University of Utah, University of Virginia, University of Washington, University of Wisconsin, Vanderbilt University, and Yale University.

\end{acknowledgements}


\bibliographystyle{aa} 
\bibliography{aa} 

\end{document}